%% file: ms.tex
\pdfoutput=1 

\documentclass[a4paper, 11pt, final]{article}

\input{ms_a_preamble}

\begin{document}

%TC:ignore

% more Hyperref options (must be after \begin{document}
\renewcommand{\figureautorefname}{Fig.}
\onehalfspacing

%--------------------------------------------------------%
%	TITLE PAGE
%--------------------------------------------------------%

\input{ms_b_frontmatter}

%TC:endignore

%--------------------------------------------------------%
%	CONTENT
%--------------------------------------------------------%

\newpage

% {
%     \setlength{\parindent}{0cm}
%   \begin{abstract}
%     A novel procedure is presented for the objective comparison and evaluation of a bank's decision rules in optimising the timing of loan recovery. This procedure is based on finding a delinquency threshold at which the financial loss of a loan portfolio (or segment therein) is minimised. Our procedure is an expert system that incorporates the time value of money, costs, and the fundamental trade-off between accumulating arrears versus forsaking future interest revenue. Moreover, the procedure can be used with different delinquency measures (other than payments in arrears), thereby allowing an indirect comparison of these measures. We demonstrate the system across a range of credit risk scenarios and portfolio compositions. The computational results show that threshold optima can exist across all reasonable values of both the payment probability (default risk) and the loss rate (loan collateral). In addition, the procedure reacts positively to portfolios afflicted by either systematic defaults (such as during an economic downturn) or episodic delinquency (i.e., cycles of curing and re-defaulting). In optimising a portfolio's recovery decision, our procedure can better inform the quantitative aspects of a bank's collection policy than relying on arbitrary discretion alone.
%     \end{abstract}
    
%      % Insert keywords here
%     \keywords{Optimisation; Credit Loss; Loan Delinquency; Collections; Expert systems}
     
%      % Insert JEL codes here
%      \JEL{C44, C63, G21.}
% }

\input{1-Intro}
\input{2-Background}
\input{3-Method}

\input{4-Results}

\input{5-Conclusion}

\subsection*{Acknowledgements}
\noindent This study is financially supported by the Absa Chair in Actuarial Science, hosted at the University of Pretoria, with no known conflicts of interest that may have influenced the outcome of this work. The authors would like to thank all anonymous referees and editors for their valuable contributions that improved this work.

%TC:ignore

\input{6-Appendix}

%--------------------------------------------------------%
%	REFERENCE LIST
%--------------------------------------------------------%

\singlespacing
\printbibliography % using biblatex
%\section*{References}
%\bibliographystyle{newapa}
% see http://texdoc.net/texmf-dist/doc/latex/natbib/natbib.pdf for more styles
%\bibliography{bibliography} 
\onehalfspacing

%TC:endignore

%--------------------------------------------------------%
%	END OF DOCUMENT
%--------------------------------------------------------%

\end{document}

%% file: ms_a_preamble.tex
\usepackage{amssymb} % provides various useful mathematical symbols
\usepackage{amsthm} % provides extended theorem environments
\usepackage{newtxmath} % provides additional math symbol support in Times New Roman
\usepackage{amsmath,empheq}

\DeclareMathAlphabet{\mathcal}{OMS}{cmsy}{m}{n}
\DeclareMathAlphabet\mathbfcal{OMS}{cmsy}{b}{n}

\DeclareFontFamily{U}{dutchcal}{\skewchar\font=45 }
\DeclareFontShape{U}{dutchcal}{m}{n}{<-> s*[1.0] dutchcal-r}{}
\DeclareFontShape{U}{dutchcal}{b}{n}{<-> s*[1.0] dutchcal-b}{}
\DeclareMathAlphabet{\mathcald}{U}{dutchcal}{m}{n}
\SetMathAlphabet{\mathcald}{bold}{U}{dutchcal}{b}{n}
\DeclareMathAlphabet\mathcalz{T1}{pzc}{mb}{it}

\DeclarePairedDelimiter\floor{\lfloor}{\rfloor}

\usepackage[utf8]{inputenc}
\usepackage{microtype} % Slightly tweak font spacing for aesthetics
\usepackage{newtxtext} % change font to Adobe Times New Roman
\usepackage[british]{babel}
\usepackage{csquotes}		%Necessary for biber

\providecommand{\JEL}[1]{\textit{\textbf{JEL: }} #1}
\providecommand{\keywords}[1]{\textbf{\textit{Keywords--- }} #1}

\usepackage{setspace} 

\usepackage{parskip} % each new line automatically spaces previously paragraph correctly
\usepackage{etoolbox}
\usepackage{titlesec}
\titleformat{\section}{\normalfont\Large\bfseries}{\thesection}{1em}{}
\titleformat{\subsection}{\normalfont\large\bfseries}{\thesubsection.}{1em}{}
\titleformat{\subsubsection}{\normalfont\normalsize\itshape}{\thesubsubsection.}{1em}{}

\usepackage{abstract}
\renewenvironment{abstract}
 {\normalfont
  \begin{center}
  \bfseries \abstractname\vspace{-.5em}\vspace{0pt}
  \end{center}
  \list{}{
    \setlength{\leftmargin}{0cm}%
    \setlength{\rightmargin}{\leftmargin}%
  }%
  \item\relax}
 {\endlist}

\usepackage{authblk} % package for author affiliations
\usepackage[bottom]{footmisc} % Makes footnotes stick to bottom of the page

\usepackage{graphicx} % More advanced figure inclusion
\usepackage{float} % For specifying table/figure locations, i.e. [ht!]
\usepackage{subcaption}
\usepackage{afterpage} % For encapsulating a floating figure on a single page   

\usepackage{printlen}

\usepackage[labelfont=bf]{caption}
\usepackage{color, colortbl}
\definecolor{LightGray}{rgb}{0.93,0.914,0.914}    
\usepackage{longtable,rotating} % For long tables that span multiple pages
\usepackage{booktabs} % used for making more professional-looking tables
\usepackage{multirow} % used for cell-merging in complex tables
\usepackage{arydshln} % for drawing other line types in tables
\PassOptionsToPackage{hyphens}{url} % for hyphenating URLs in main text

\newcommand*\rot{\rotatebox{90}}

\usepackage{fancyhdr}
\fancyheadoffset{0cm}
\makeatletter
\newcommand{\quickwordcount}[1]{
  \immediate\write18{texcount -quiet -incbib -sub=none -utf8 -1 -sum -merge -encoding=utf8 #1.tex > #1-words}%
  \immediate\openin\somefile=#1-words
  \read\somefile to \@@localdummy
  \immediate\closein\somefile
  \setcounter{wordcounter}{\@@localdummy}
  \@@localdummy
}
\makeatother

\usepackage{silence}
\usepackage[style=apa,backend=biber,natbib,hyperref]{biblatex}
\DeclareLanguageMapping{british}{british-apa}
\defbibenvironment{bibliography}{\enumerate}{\endenumerate}{\item}

\usepackage[colorlinks=false,allcolors=black]{hyperref} 
\let\orgautoref\autoref

\renewcommand{\autoref}[1]
{%
\def\equationautorefname{Eq.}%
\def\figureautorefname{Fig.}%
\def\subfigureautorefname{Fig.}%
\orgautoref{#1}%
}

\usepackage[T1]{fontenc}

\usepackage[nameinlink,capitalise]{cleveref}
\usepackage{algorithm,algpseudocode}

\makeatletter
\newlength{\trianglerightwidth}
\algnewcommand{\LineCommentCont}[1]{\Statex \hskip\ALG@thistlm%
  \parbox[t]{\dimexpr\linewidth-\ALG@thistlm}
{\leftskip=\algorithmicindent
  \hangindent=\algorithmicindent 
  \hangafter=1%
  \strut\makebox[\algorithmicindent][c]{$\triangleright$}#1\strut}
  } % \trianglerightwidth
\makeatother

\usepackage[left=1.8cm, right=1.8cm, bottom=2.5cm, top=2.5cm]{geometry}

%% file: ms_b_frontmatter.tex
%--------------------------------------------------------%
%	TITLE
%--------------------------------------------------------%

% Article title
\newcommand{\MainTitleText}{Simulation-based optimisation of the timing of loan recovery across different portfolios}

\title{\fontsize{20pt}{0pt}\selectfont\textbf{\MainTitleText
}}

%--------------------------------------------------------%
%	AUTHORS
%--------------------------------------------------------%   

\author[,a]{\large Arno Botha \thanks{ \url{arno.spasie.botha@gmail.com}; ORC iD: 0000-0002-1708-0153}}
%\author[,a]{\large Arno Botha \thanks{ \url{arno.spasie.botha@gmail.com}}}
%\author[,a]{\large Conrad Beyers \thanks{Corresponding author: \url{conrad.beyers@up.ac.za}}}
\author[,a]{\large Conrad Beyers \thanks{ \url{conrad.beyers@up.ac.za}}}
%\author[,b]{\large Pieter de Villiers}
\author[,b]{\large Pieter de Villiers\thanks{\url{pieter.devilliers@up.ac.za}}}
\affil[a]{\footnotesize \textit{Department of Actuarial Science, University of Pretoria, Private Bag X20, Hatfield, 0028, South Africa}}
\affil[b]{\footnotesize \textit{Department of Electrical, Electronic, and Computer Engineering, University of Pretoria, Private Bag X20, Hatfield, 0028, South Africa}}
\renewcommand\Authands{, and }

% Today's date
 	%\date{Submitted: \usvardate\today}
    
%by specifying the below, we essentially "rewrite" the command \maketitle, which is normally called in main.tex.
%this is done primarily to abuse the \date command above

\makeatletter
\renewcommand{\@maketitle}{
    \newpage
     \null
     \vskip 1em%
     \begin{center}%
      {\LARGE \@title \par
      	\@author \par}
     \end{center}%
     \par
 } 
 \makeatother
 
 \maketitle

%--------------------------------------------------------%
%	ABSTRACT
%--------------------------------------------------------%    
{
    \setlength{\parindent}{0cm}
    \rule{1\columnwidth}{0.4pt}
    \begin{abstract}
    A novel procedure is presented for the objective comparison and evaluation of a bank's decision rules in optimising the timing of loan recovery. This procedure is based on finding a delinquency threshold at which the financial loss of a loan portfolio (or segment therein) is minimised. Our procedure is an expert system that incorporates the time value of money, costs, and the fundamental trade-off between accumulating arrears versus forsaking future interest revenue. Moreover, the procedure can be used with different delinquency measures (other than payments in arrears), thereby allowing an indirect comparison of these measures. We demonstrate the system across a range of credit risk scenarios and portfolio compositions. The computational results show that threshold optima can exist across all reasonable values of both the payment probability (default risk) and the loss rate (loan collateral). In addition, the procedure reacts positively to portfolios afflicted by either systematic defaults (such as during an economic downturn) or episodic delinquency (i.e., cycles of curing and re-defaulting). In optimising a portfolio's recovery decision, our procedure can better inform the quantitative aspects of a bank's collection policy than relying on arbitrary discretion alone.
    \end{abstract}
    
     % Insert keywords here
    \keywords{Optimisation; Credit Loss; Loan Delinquency; Collections; Expert systems}
     
     % Insert JEL codes here
     \JEL{C44, C63, G21.}
    
    \rule{1\columnwidth}{0.4pt}
}

\noindent Word count (excluding front matter and appendix): 7662 %\quickwordcount{ms} 

\noindent Figure count: 9

%% file: 1-Intro.tex
\section{Introduction}
\label{sec:ch1}

Consumer credit has exponentially grown over the last few decades, largely spurred by the introduction of the credit card during the 1950s. Its current estimate of approximately \$11 trillion from the US market consists largely of mortgages, credit cards, personal loans, vehicle financing, overdrafts and other revolving loans for the individual, as reported in \citet{fred2020}. This credit growth, as argued in \citet[pp.~1--6]{thomas2009book} and \citet{thomas2010consumer}, could not have been possible without a degree of automation, historically facilitated by statistical decision-making models otherwise known as credit scorecards. These models rendered consistent approve/decline credit decisions that enabled greater application volumes whilst keeping default risk under control, i.e., the risk of the borrower reneging on repayments. This control is mainly achieved by only approving those applications with a predicted probability of default within a desired limit, which is usually aligned with a bank's risk appetite. Constructing these scorecards involves finding a statistical relationship between a set of borrower-specific characteristics and the successful (or failed) repayment outcome over time, using historical data. Naturally, the credit scoring literature on optimising this relationship is considerable, including various machine learning approaches; see \citet{hand1997}, \citet{hand2001m}, \citet{thomas2002book}, \citet{siddiqi2005credit}, \citet{crook2007recent}, \citet{thomas2009book}, \citet{thomas2010consumer}, \citet{hao2010review}, and \citet{louzada2016classification}. However, the vast majority hereof assumes a preset outcome definition, with little effort spent on improving the definition itself.

The advent of these automated models did, however, call for a more methodical manner of "measuring default" before trying to predict the risk thereof. In most cases, the development of loan delinquency over time is captured using the number of payments in arrears from accountancy practices, which is constructed from days past due (DPD). Whilst practical and intuitive, this calculation (or the $g_0$ delinquency measure as we shall call it) has a few flaws upon which alternative measures may improve, as discussed in the appendix. Nonetheless, banks commonly specified three payments (or 90 DPD) in arrears as a pragmatic point of `default', long before the introduction of the Basel II Capital Accords. That said, this threshold can generally range between 30--180 days based on managerial discretion and some analysis, as discussed in \citet[pp.~123--124]{thomas2002book} and later in \autoref{sec:ch2}. However, the direct financial implications of any chosen definition are not readily known nor accounted for when deciding the point of default during typical analyses, especially when developing credit scoring models. Therefore, and as originally argued in \citet{hand2001m}, pursuing modelling excellence becomes questionable when the constructed outcome variable, itself determined by the default definition, is inherently quite arbitrary.

\begin{figure}[ht!]
	\centering
	\includegraphics[width=0.95\linewidth,height=0.4\textheight]{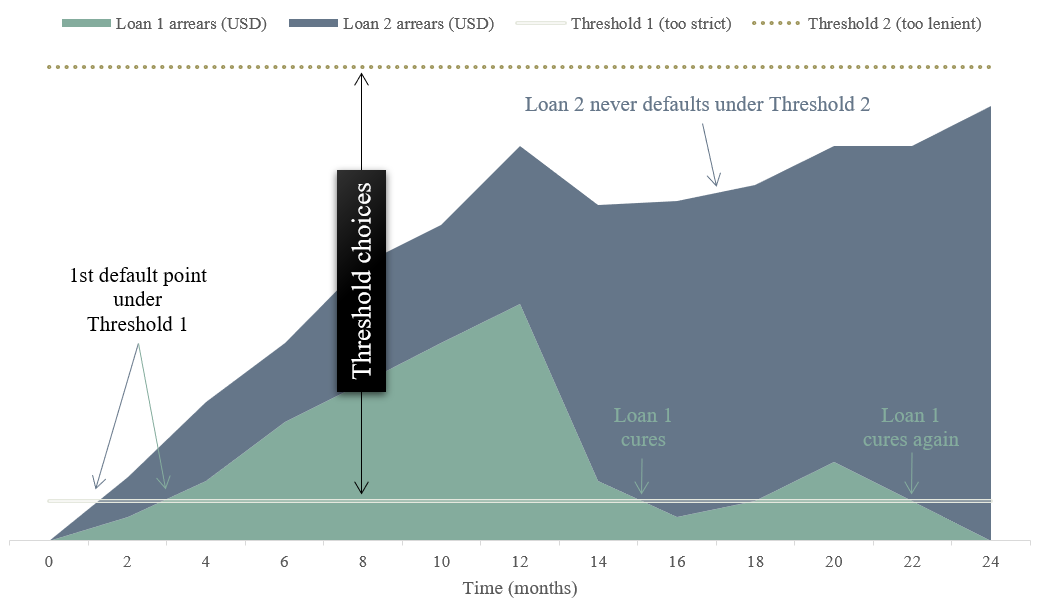}
	\caption{Illustrating the trade-off associated with two extreme arrears-based thresholds for two fictional loans of the same size. Threshold 1 is overly strict for loan 1 given that it cures later; but suitable for loan 2 since it never cures. Conversely, threshold 2 is overly naive for loan 2, though suitable for loan 1.}
	\label{fig:trade-offs}
\end{figure}

Fundamentally, as an account continues to accrue arrears, the lender will respond by proportionately ramping up its collection efforts. Every unpaid instalment (or portion thereof) erodes the trust between bank and borrower, which is only tolerable up to a point. This ambiguous point may itself differ across portfolios and even banks, likely based on differences in risk appetites and market conditions. Regardless, having reached this point, the bank effectively assumes that the troubled loan will helplessly fall into ever greater arrears if kept. Therefore, the lender shifts its focus to the immediate and maximal recovery of debt, including selling any collateral, as based on the five-phase credit management model of \citet[pp.~11--13,~147--153]{finlay2010book}. Presumably, this idea of reaching a so-called "point of no return" is the historical basis for a default definition, although most modern definitions also contain more qualitative criteria. In addition, a loan may `cure' from default whenever a borrower repays the arrears (regardless of reason), which further casts doubt on a chosen default threshold as the supposed "point of no return". For Basel-compliant or IFRS 9-compliant types of credit risk modelling, another factor to consider is that of competing regulatory requirements when defining `default' across different jurisdictions. Furthermore, some lenders use multiple definitions for different purposes or across different portfolios -- all of which impedes the interpretation of `default' in trying to cater for so many different contexts.

Owing to the difficulties of defining `default' precisely, we explore a more fundamental meaning of `default' as the portfolio-dependent, probabilistic, and risk-based "point of no return" beyond which loan collection becomes sub-optimal if pursued. Our `default' state is simply based on breaching a certain delinquency threshold, so that the "net cost" of each candidate threshold can be assessed. The \textit{best} time is sought at which the lender should forsake a loan and instead try to collect all it can. Furthermore, it is convenient to try and find this point from a loan delinquency-basis since the resulting measurements are scale-invariant and already incorporate behavioural information on the borrower. Too strict a delinquency threshold will surely marginalise accounts that would have resumed repayment (or cured from `default'), had the bank been more patient before initiating strict recovery. A loan may also experience multiple episodes of `redefaulting' and curing, which is further exacerbated by a threshold that is too strict. Conversely, too lenient a threshold may naively tolerate increasing arrears at the cost of greater liquidity risk and bigger capital buffers, which may introduce capital-inefficiencies. The goal now becomes to devise an expert system in which these two extremes can be reasonably offset against each other. Doing so can theoretically form a proverbial `Goldilocks-zone' in space that contains the ideal delinquency threshold for a portfolio, which translates to the `best' time for loan recovery. This concept is illustrated in \autoref{fig:trade-offs} using the arrears amount (proportional) as a high-level threshold, including two extreme choices thereof.

\begin{figure}[ht!]
	\centering
	\includegraphics[width=0.65\linewidth,height=0.1\textheight]{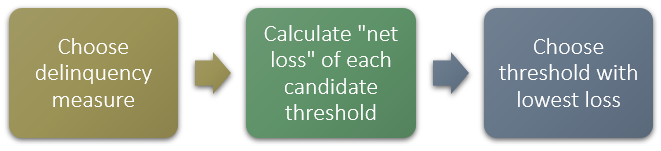}
	\caption{High-level steps of the contributed LROD-procedure.}
	\label{fig:steps}
\end{figure}

In this study, we develop such a system, called the \textit{Loss-based Recovery Optimisation across Delinquency} (LROD) procedure, as our main contribution. This procedure is summarised in three steps, shown in \autoref{fig:steps} and formally presented in \autoref{sec:loss_proc}. Relevant literature is explored in \autoref{sec:ch2}, including current practices on selecting default definitions as well as previous optimisation work on loan collection. Since different portfolios will likely have different `ideal' recovery thresholds, a simple simulation-based setup is described in \autoref{sec:sim_techs} as our testbed. This allows for examining recovery optimisation from "first principles" by randomly generating amortising loan portfolios with specifiable risk profiles, guided by expert judgement and industry experience. Moreover, by tweaking this testbed's simulation parameters appropriately, one can obtain quick managerial insight on the viability of recovery optimisation before embarking on any deep data work; a useful secondary contribution. We demonstrate the LROD-procedure in \autoref{sec:ch4} by conducting a broad computational study using the aforementioned testbed. Threshold optima are found across most levels of default and loss risk, as measured by the probability of payment and loss rate respectively. Furthermore, portfolios suffering from systematic pattern-like defaults are tested, as well as portfolios with episodic delinquency (cycles of curing and re-defaulting). Overall, our procedure delivers dynamic yet intuitive optimisation results for the timing of loan recovery. It may therefore be used to improve a bank's existing collection policies, with the accompanying source code published in \citet{botha2020sourcecode1}. Finally, we conclude the study in \autoref{sec:ch5} and outline areas of future research.

%% file: 2-Background.tex
\section{An overview of loan default and collections optimisation}
\label{sec:ch2}

The estimation of the frequency of any event in a given sample fundamentally depends on the definition of the event. While loan `default' lies intrinsic to credit risk (and its estimation), the phenomenon thereof certainly has many definitions, both historically and in modern times. These definitions typically vary by product, customer type, and bank, as discussed in \citet[pp.~203--212]{VanGestel2009book} and \citet[pp.~137--138]{baesens2016credit}. Examples hereof include filing for bankruptcy, unfulfilled claims, negative net present values, overdrawing beyond an agreed credit limit, as well as becoming three instalments in arrears. Basel II standardised default definitions to some extent upon its introduction, while still leaving room for the lender's discretion, subject to regulatory approval. Specifically, paragraph 452 of the \citet{basel2006} defines `default' as one of the following two conditions. Firstly, the obligor has reached 90 DPD on a material loan balance, or has been in excess of an advised credit limit for 90 days. Alternatively, the bank considers, \textit{in its opinion}, that the obligor is unlikely to repay its obligations in full, without the necessary intervention of the bank, e.g., liquidating any collateral. To help inform this opinion, Basel II includes a few reasonable (but qualitative) indicators of \textit{"unlikeliness to pay"} in paragraph 453. Examples include when debt restructuring leads to an overall reduced obligation, or partially selling off a debt at a loss.

The requirements of Basel II are often promulgated \textit{verbatim} by some regulators, e.g., Regulation 67 of the Banks Act of \citet[pp.~1201--1202]{banksact1990}. However, Basel II (and how it relates to `default') is only truly relevant to estimating the amount of capital required for offsetting \textit{unexpected losses} (UL). In turn, this requires modelling the \textit{expected losses} (EL) for which a bank also holds provisions. EL is generally defined as the product of three specific risk parameters: 1) the Probability of Default (PD); 2) the Loss Given Default (LGD); and 3) the Exposure-At-Default (EAD). A comprehensive review of this topic is given in \citet[pp.~289--293]{thomas2009book}, \citet[chap.~4,~6]{VanGestel2009book}, and \citet[chap.~5--11]{baesens2016credit}. Relatedly, the new accounting standard \citet{ifrs9_2014} firmly lodged the management of loss provisions as a deeply statistical exercise similar to that of capital estimation, which is discussed in \citet{novotny2016}, \citet{skoglund2017} and \citet{cohen2017new}. However, even IFRS 9 does not impose a fixed default definition, instead requiring in paragraph B.5.5.37 that a particular definition simply be used consistently in a portfolio's risk management. In addition, IFRS 9 presumes 90 DPD as a default definition that may be superseded by alternatives if they are demonstrated as \textit{"reasonable"}.

Basel II and IFRS 9 regulate certain aspects of a default definition as it pertains to specific exercises, i.e., modelling the UL and EL. However, the notion of `default' extends to other areas in retail banking as well, most notably that of credit scoring, pricing, and collections. Selecting appropriate default definitions within these areas are often based on managerial discretion though supported by some analysis. In particular, the observed transitions amongst increasingly severe arrears categories (30 days, 60 days, etc.) are cross-tabulated across a chosen length of time in what is called a \textit{roll rate analysis}. From \citet[pp.~33--42]{siddiqi2005credit}, the principle is to select a particular category as the default definition that is sufficiently stable in that accounts identified as `lost' ought to remain lost at the end of the outcome period. The chosen category should yield a minimum of accounts recovering from `default' on average. However, the direct loss implications associated with any definition may be a better criterion than stability since the latter ignores any competing financial/opportunity costs that may actually exist when varying the default threshold. Furthermore, the `true' transition rates can be obscured by the epoch of time from which loan performance is sampled, which can certainly influence the chosen threshold. Perhaps the greatest source of variation underlying these roll rates is the length of the outcome period, which can vary between 6--24 months in practice, as discussed in \citet[pp.~99]{thomas2002book} and \citet[pp.~101--102]{VanGestel2009book}. 

The work of \citet{kennedy2013window} and \citet{mushava2018experimental} investigated the role of the outcome period using Irish and South African data respectively. The authors used different time spans in predicting default risk and found that too short a window becomes insufficient in capturing the transition rates due to seasonal effects and/or risk immaturity. Conversely, overly long windows may no longer represent the portfolio's current risk composition, strategies, or even the current market conditions, in addition to yielding models with degrading accuracy. Furthermore, longer windows can ignore rapid transitions amongst delinquency states, e.g., oscillating between defaulting and curing, as discussed in \citet{kelly2016good}, which is especially relevant for the accuracy of monthly EL estimates. The outcome length and the sample window are clearly significant factors that complicate the choice of a default definition. As an example, a particularly low curing rate given a chosen definition cannot truly justify the latter (without conducting additional analysis) due to these other confounding factors. Put differently, low curing rates may instead be attributed to an overly short outcome period or shifting market conditions -- both of which are reasons why a roll rate-based approach is deemed unfit for dynamically finding the "point of no return" in this study.

Varying the default threshold within a definition was first studied in \citet{harris2013quantitative} and \citet{harris2013default} from a credit scoring perspective. The authors built default-classifiers using Support Vector Machines across various thresholds and found that the model accuracy is affected by the chosen threshold. However, while optimising accuracy is certainly worthwhile, these results say little about the direct impact on profitability when varying the default threshold. As originally argued in \citet{hand1997} and \citet{hand2001m}, a lender is primarily interested in the underlying profitability of a credit decision, with credit risk being but a facet thereof. Surely, borrowers with no arrears are likely to be profitable ventures for the bank, while accruing arrears up to a point can certainly lead to eventual losses. However, there is little objective evidence in literature for justifying the presumption of profitability underlying 90 DPD as the \textit{ideal} default threshold. Moreover, not all `defaults' (or default thresholds) are equal, as demonstrated in \citet{kelly2016defaults} using Irish mortgage data. A legal peculiarity during 2009--2013 made it extremely difficult for Irish lenders to liquidate troubled mortgages, which led to disproportionately deep levels of arrears. The authors modelled so-called `deep defaults' (e.g., 360+ DPD) across different arrears severities that were then used as default thresholds, which yielded markedly different curing experiences. Amongst other things, these results cast doubt on the supposed finality of the classical 90 DPD threshold serving as the "point of no return".

Selecting any DPD-based threshold will affect the associated probability of curing from the supposed `default' state. However, the chosen threshold's suitability as a "point of no return" becomes questionable whenever this probability is nonzero. Furthermore, multi-period "episodes of delinquency" are more widespread in practice than one would otherwise believe, based on anecdotal experience. The work of \citet{thomas2016} demonstrated these patterns of periodic repayments using a four-state homogeneous Markov chain, in modelling the collections process of defaulted UK loans. The authors noted that these models may be used to evaluate write-off policies, even though this is not truly recommended. Instead, pursuing loan collection for an optimal length of time was investigated in \citet{mitchner1957operations}, based on maximising net profit using US loans. The authors found that loan recovery should cease whenever the one-period expected repayment equals the collection cost itself. However, they assumed that a defaulted borrower is permanently absorbed into a paying regime once cured, which contrasts \citet{thomas2016}. Regardless, Markov chains have long been used in modelling credit risk (or aspects thereof), thereby motivating its use in this study. See \citet{cyert1962} for a Markovian approach in estimating the allowance for bad debts, later extended in \citet{corcoran1978use} and \citet{jarrow1997markov}, and reviewed more recently in \citet{hao2010review}.

A dynamic programming model was formulated in \citet{de2010optimizing} in optimising the collections process using unsecured European loans. The idea is to find the ideal recovery action and its optimal pursuit duration, which maximises the net recovery rate for an "average" debtor. These actions include telephonic calls, demand letters, house visits, threats, legal steps, and write-off. However, cash flows from previous or future periods were excluded from the state space formulation, which limits the approach's tractability. This work was extended in \citet{so2019debtor} by following a Bayesian approach on the individual debtor-level to give similarly optimised outputs. Within the same problem context, a Markov decision process was developed in \citet{liu2019markov} as an alternative approach using designed data. Similarly, an optimal collection action is sought across both delinquency states and time. The authors calculated a schedule of optimised actions based on maximising expected net present value, which superseded a static collection policy as an alternative. However, strong assumptions were made when designing both the data and elements within the authors' method, which may not be suitable in practice. Moreover, write-off was not structured as a candidate collection action, instead being exogenously imposed within the Markov chain's state space.

In form, our study is closest to that of \citet{de2010optimizing} and \citet{liu2019markov}, though a different and arguably more general approach is followed. Specifically, we use delinquency measures instead of time and leverage the entire portfolio instead of using only `defaulted' loans, which already imposes a particular "point of no return". We focus more fundamentally on if and when to abandon a loan based on accrued delinquency, instead of pursuing various collection actions. From a literature perspective, our work attempts to bridge the branches of credit risk modelling and collection optimisation. This is achieved by framing the recovery decision's timing as a loss-based optimisation problem wherein the \textit{ideal} "point of no return" is sought.

%% file: 3-Method.tex
\section{An approach for optimising and testing the recovery decision}
\label{sec:ch3}

\begin{figure}[ht!]
\centering\includegraphics[width=0.7\linewidth,height=0.37\textheight]{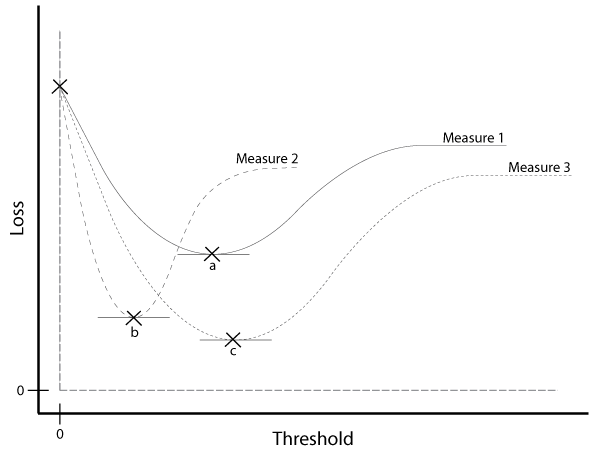}
\caption{Illustrating the loss optimisation of the recovery decision across several delinquency measures. As a result, Measure 3 is chosen as the best measure with its minimum loss attained at threshold $c$.}\label{fig:approach1}
\end{figure}

\textit{Delinquency} is defined as a time-dependent, varying, and measurable quantity that represents the extent of eroded trust between bank and borrower. Relatedly, a \textit{delinquency measure} $g$ reflects the degree of non-payment based on the fundamental idea of a borrower owing an amount $I_t>0$ (instalment) though only repaying an amount $R_t\geq0$ (receipt) at a particular time $t$. When $R_t < I_t$, the function $g$ quantifies the extent $I_t - R_t$ by which the bank incrementally loses confidence in the borrower honouring the original credit agreement. Three different delinquency measures $g_1$, $g_2$, and $g_3$ are used in this study, with their construction detailed in the appendix. We develop the so-called \textit{Loss-based Recovery Optimisation across Delinquency} (LROD) procedure in \autoref{sec:loss_proc}. This procedure attempts to find the `best' delinquency-based threshold for a chosen measure $g\in \big\{g_1,g_2,g_3 \big\}$ at which the portfolio's recovery decision is loss-optimised, as illustrated in \autoref{fig:approach1}. In addition, a simulation-based setup is described in \autoref{sec:sim_techs} by which portfolios can be systematically generated across various credit risk scenarios, in testing the LROD-procedure.

\subsection{Optimising loan recovery times: the LROD-procedure}
\label{sec:loss_proc}

Consider a portfolio of $N$ loans, indexed by $i=1,\dots,N$, and let $g(i,t)$ denote the value of a particular measure $g\in \big\{g_1,g_2,g_3 \big\}$ at periods $t=0,\dots,t_{c_i}$ with $t_{c_i}$ representing the contractual term of the $i^\text{th}$ account. Let $v_t^{(a)}$ and $v_t^{(b)}$ be standard discounting functions that use an alternative risk-free interest rate and the loan interest rate respectively in discounting back $t$ periods, both expressed as annual effective rates and parametrised later. Let $R_t^i$ and $I_t^i$ be the respective receipt and expected instalment at time $t$ for the $i^\text{th}$ account. Then, let $R(i,t)$ be the summed historical receipts up to $t$, expressed as \begin{equation} \label{eq:discounted_receipts}
	R(i,t) = \sum_{l=0}^{t}{R_l^iv_l^{(a)}} \ .
\end{equation} For the remaining future instalments, let $O(i,t)$ denote the expected outstanding balance at $t$, defined as \begin{equation} \label{eq:expectedbalance}
	O(i,t) = v_t^{(a)} \sum_{l=t+1}^{t_{c_i}} {I_l^i v_{l-t}^{(b)}}, \quad O(i,t) = 0 \quad \text{for} \ t=t_{c_i} \ .
\end{equation} To cater for arrears, let $A(i,t)$ be the historical and cumulative shortfall up to $t$ between instalments and receipts, given by \begin{equation} \label{eq:arrears}
	A(i,t) = \sum_{l=0}^{t} {\left(I_l^i - R_l^i \right)v_l^{(a)}} \ .
\end{equation}

Financial loss can only be realised when the lender disposes of the impaired asset, regardless of the extent of impairment. In this study, `default' is interpreted as a variable state, which will become useful for optimising the eventual recovery decision. Having breached some threshold (signifying broken trust), the lender's objective changes to collecting the maximum in the shortest time possible. As a simplifying assumption, a fixed portion of the loan is immediately written-off upon entering `default'. In reality, this portion will likely depend on many factors, including the workout period itself. This assumption can certainly be relaxed in future research when refining this optimisation procedure and what is essentially its LGD-component. Accordingly, let $r_E \in[0,1]$ be a loss rate applied on $O(i,t)$ to  help reflect any underlying opportunity costs of forsaking future revenue. Moreover, assume that $A(i,t)$ is partly written-off at a different loss rate $r_A \in[0,1]$ to account for impairment. Using two different rates recognises that the recovery success may differ between these two components (expected balance and arrears). Secondly, this setup accounts for implicit trade-offs between forsaking future revenue versus accruing arrears for a given $t$ respective to \autoref{eq:expectedbalance} and \autoref{eq:arrears}. These unconditional loss rates serve as placeholders for more sophisticated loss models (or expert knowledge of the loss experience), presumably including all other costs. Finally, let $l(i,t)$ be the discounted "blended loss" assessed at $t$ and expressed as \begin{equation} \label{eq:discountedloss}
	l(i,t) = O(i,t)r_E + A(i,t)r_A \ .
\end{equation} 

For optimising loan recovery, let $d\geq0$ be a delinquency threshold such that the $i^\text{th}$ account is considered as $(g,d)$-defaulting if and only if $g(i,t)\geq d$ at any particular time $t=1,\dots,t_{c_i}$. Let $\mathcal{S}_D$ be the subset of all $(g,d)$-defaulting accounts such that \begin{equation} \label{eq:subset_d}
	\mathcal{S}_D = \big\{ i \ \big| \ \exists  \ t \in [0,t_{c_i}] \ : \ g(i,t) \geq d \big\} \ .
\end{equation} Since an account may enter and leave the $(g,d)$-default state multiple times in reality, let $t_i^{(g,d)}$ be the earliest moment of `default' for a qualifying account, defined as \begin{equation} \label{eq:earliest_default_t}
	t_i^{(g,d)} = \min { \big(t : g(i,t) \geq d \big)}, \quad \forall \ i \in \ \mathcal{S}_D \ .
\end{equation} Similarly, let $\mathcal{S}_P$ be the subset of all accounts considered as $(g,d)$-performing such that \begin{equation} \label{eq:subset_p}
	\mathcal{S}_P = \big\{i : \ g(i,t) < d \quad \forall \ t \in [0,t_{c_i}] \big\} \ .
\end{equation} The difference in assessing losses between a $(g,d)$-defaulting and a $(g,d)$-performing account is simply the time of assessment $t$, which is set at either $t=t_i^{(g,d)}$ or $t=t_{c_i}$ respectively within $l(i,t)$ from \autoref{eq:discountedloss}. At each time $t$, the lender effectively decides an account's membership between $\mathcal{S}_D$ or $\mathcal{S}_P$, based on accrued delinquency $g(i,t)$ and a particular $(g,d)$-configuration. The latter is to be adopted as a portfolio-wide delinquency-based collection policy at the outset $t=0$, as is common South African collection practice. In a sense, accrued delinquency forms the time-invariant \textit{action} space of a Markov decision process (MDP) in choosing $d$, whereas accrued delinquency formed the \textit{state} space in \citet{liu2019markov}. Accordingly, our state space is set membership itself, i.e., either $\mathcal{S}_P$ or $\mathcal{S}_D$. However, we do not employ a classical MDP framework, instead opting for a simpler approach that facilitates choosing a static $(g,d)$-policy. Both `payoff' and the element of time is already accounted for in \autoref{eq:discountedloss} by having discounted the associated loss to $t=0$ for a given $(g,d)$-policy. As such, the objective function is simply the total portfolio loss $L_g(d)$ for a particular $(g,d)$-configuration, defined as \begin{equation} \label{eq:total_loss}
	L_g(d) = \sum_{i \in \ \mathcal{S}_D} {l\left(i,t_i^{(g,d)} \right)} \ + \ \sum_{i \in \ \mathcal{S}_P} {l\left(i,t_{c_i}\right)} \ .
\end{equation}

Losses are iteratively calculated across a range of thresholds $d \in \mathcal{D}_g$ using $L_g$ from \autoref{eq:total_loss} with a particular measure $g\in \big\{g_1,g_2,g_3 \big\}$. In summary, three preparatory steps are necessary before conducting optimisation: \begin{enumerate}
	\item Delinquency must be measured for every account and across its history using $g\in \big\{g_1,g_2,g_3 \big\}$;
	\item Select appropriate thresholds $d \in \mathcal{D}_g$ on the domain of a particular $g$ for optimisation;
	\item A portfolio loss model $L_g$ must be applied for every chosen threshold $d \in \mathcal{D}_g$ of each $g$.
\end{enumerate} The main recovery optimisation problem is effectively divided into smaller $(g,d)$-based sub-problems. The resulting $L_g(d)$ for each $(g,d)$-configuration is stored centrally, thereby forming a loss curve across $d$ for each $g$. The objective then becomes a search for a threshold $d'\in\mathcal{D}_g$ such that $L_g(d')\leq L_g(d)$ for all chosen $d\in\mathcal{D}_g$. More generally, if a global minimum $m^{(g)}$ exists on $L_g$ for a particular measure $g$, then $L_g$ is said to be minimised at $d^{(g)}$. Minimising again across the set formed by $m^{(g)}$ effectively allows indirect comparison of delinquency measures themselves at the portfolio-level. The optimal measure $g^{\ast}$ is then the $g$ that yielded the lowest loss at its corresponding threshold, as illustrated in \autoref{fig:approach1} and expressed as \begin{equation} \label{eq:loss_proc_g}
    g^{\ast}=\arg_g\,\min_{g\in \left\{g_1,g_2,g_3 \right\}}{\left[m^{(g_1)}, m^{(g_2)}, m^{(g_3)}\right]} \, .
\end{equation}

Alternatively, a single measure can be used, e.g., $g_1$, which simplifies the optimisation to finding a threshold $d^{(g_1)} \in \mathcal{D}_{g_1}$ that equals $\arg_d\,\min{ L_{g_1}(d)}$ if a minimum exists. Nonetheless, the optimisation's feasibility relies heavily on adequately populating the search space $\mathcal{D}_g$. Thresholds are trivially chosen as $d=0,\dots,d_N$ for the integer-valued $g_1$-measure since $\mathcal{D}_{g_1}$ is a countable set, where $d_N$ is a reasonable (but admittedly arbitrary) proportion of the maximum contractual term, e.g., $60\%$. However, this becomes more complicated for the real-valued measures $g_2$ and $g_3$ since their search spaces contain infinite possible thresholds. As such, two competing interests are balanced against each other when assembling $\mathcal{D}_g$: 1) too few thresholds that are inadequately spaced can obscure hidden optima and ruin the optimisation; 2) too many thresholds can become computationally burdensome. As a practical expedient, the output of $g_2$ and $g_3$ are binned into a discretionary range of thresholds by which $\mathcal{D}_g$ is populated, followed by manual tweaks.

 \subsection{Portfolio generation: a testbed for the LROD-procedure}
 \label{sec:sim_techs}
 
A real-world portfolio inherently suffers from censoring insofar that delinquent loans are only kept on the balance sheet up to a certain point, as controlled by the bank's write-off policies. Although eventually optimising the recovery decision of a real-world portfolio would be ideal, it is arguably prudent first to demonstrate the efficacy hereof from "first principles" on designed data. In this section, a broad but simple simulation-based setup is described, guided by expert judgement and industry experience. Using this setup as a testbed, replicable loan portfolios of varying risk levels are iteratively generated in testing the LROD-procedure. This testbed is subsequently used to identify a certain range of credit risk profiles for which optima are found, simply by varying the simulation parameters.

Some delinquent accounts will simply never recover in reality, which implies a continuous stream of zeros in their receipts $\boldsymbol{R}=\big[R_1,R_2,\dots,R_{t_c}\big]$ after some point. Given a measure $g\in \big\{ g_1,g_2,g_3 \big\} $ and a so-called \textit{truncation parameter} $k\geq 0$, this effect is simulated from a certain starting point $t'= \min\big(\, j : g(j) \geq k\big)$ that only exists when delinquency has accrued sufficiently, i.e., the earliest period $j\in [0,t_c]$ at which $g(j) \geq k$ is potentially triggered. A process, called $(k,g)$-truncation, then changes $\boldsymbol{R}$ to $\boldsymbol{R}'$ by \begin{equation} \label{eq:truncation}
	\boldsymbol{R}' = \begin{cases}
               \big[R_1,R_2,\dots,R_{t'},0,\dots,0\big]  & \text{if} \ t' \ \text{exists} \\
               \boldsymbol{R} & \text{otherwise}
            \end{cases} \ .
\end{equation}

Consider $N=10,000$ standard amortising loan accounts that are indexed by $i=1,\dots,N$, with a fixed contractual term of $t_c=60$ months, a fixed effective annual interest rate of 20\%, and a fixed principal amount such that the level instalment is $I_t=100$ at every period $t=1,\dots,t_c$. Admittedly, these quantities are oversimplified and will typically vary in a real portfolio based on the level of credit risk and loan demand. However, sampling them instead from stylised\footnote{In particular, a beta distribution was first parametrised to resemble the typically right-skewed distributional shape of unsecured retail loan rates in the South African market, thereby reflecting expert knowledge and risk-based pricing practises. Secondly, the loan amount was also sampled from a similarly parametrised distribution, again based on the authors' experience in the industry.} distributions did not have nearly the same effect as that of credit risk in the optimisation itself. These simplifications are therefore justified for the time being. Furthermore, an effective annual risk-free rate of 7\% is used in discounting, which is realistic for the South African market. Let the maximum loan size be $L_M=5,000$ and let $r_E=40\%$ and $r_A=70\%$ with the rationale that losses on arrears ought to be penalised more than losses on expected balances. The latter is a decreasing quantity while the former increases over time for a persistently delinquent loan. All of these parameter values represent expert knowledge though can certainly be varied in practice, which will be demonstrated later for some of these parameters.

In simulating the receipt vector $\boldsymbol{R}$ of each loan account, two probabilistic techniques are now described. As a basic technique (called \textit{random defaults}), let $u_t\in[0,1]$ be a randomly generated number at every period $t=1,\dots,t_c$ and let $b$ be the probability of payment, i.e., $\mathbb{P}(R_t=I) = b$ with $I$ denoting the level instalment. Note that $b=80\%$ is chosen as a default value, though this is later varied. Each element $R_t$ within $\boldsymbol{R}$ is then populated with either $I$ or 0, expressed as \begin{equation} \label{eq:sim_random}
	R_t = \begin{cases}
    		I & \text{if} \ u_t < b \\
            0 & \text{otherwise}
    	  \end{cases} \ .
\end{equation}

Despite its simplicity, random defaults do not feasibly generate periods of consecutive non-payments followed by resumed payment, which frequently occurs in practice as "episodic delinquency". Therefore, and similar to \citet{thomas2016}, the \textit{Markovian defaults} technique is defined where $X_t\in \{\text{P},\text{D},\text{W}\}$ denotes a random variable that can assume one of three states at each period $t$; the paying state $\text{P} : R_t=I$, the delinquent state $\text{D} : R_t = 0$, and the absorbing write-off state $\text{W} : R_{t\geq t'} = 0$ from a certain point $t'$ onwards. Then, let $X_1, X_2, \dots$ be a sequence of random variables that form a discrete-time first-order Markov chain. One can reasonably assume that every account starts off as non-delinquent, i.e., $\mathbb{P}(X_1=\text{P})=1$ while $\mathbb{P}(X_1 \in \{\text{D},\text{W}\})=0$. Subsequently, the one-period transition probability from the current state $i$ at $t$ to the future state $j$ at $t+1$ is denoted as $P_{ij}$. However, let the write-off probabilities be sensibly set to $0.1\%$ and $1\%$ respective to the starting states P and D. These values agree with general industry experience of an unsecured portfolio, though can certainly be tweaked to the individual portfolio in practice. The remaining elements in the transition matrix can now be derived from but two probabilities, $P_{\text{PP}}$ and $P_{\text{DD}}$. In turn, both of these can be systematically varied to generate a portfolio's cash flows according to a certain level (or profile) of credit risk. The transition matrix is accordingly expressed in \autoref{tab:markov_transmatrix}. \begin{table}[ht!]
\centering
\begin{tabular}{@{}ccccc@{}}
\multicolumn{1}{c}{} & \multicolumn{4}{c}{To} \\
 &  & \multicolumn{1}{c}{P} & \multicolumn{1}{c}{D} & \multicolumn{1}{c}{W}  \\ \cline{3-5}
\multirow{4}{*}{\rot{From}} & \multicolumn{1}{r|}{P} & $P_{\text{PP}}$ & $1 - P_{\text{PP}} - 0.1\%$ & \multicolumn{1}{l|}{0.1\%} \\
 & \multicolumn{1}{r|}{D} & $1 - P_{\text{DD}} - 1\%$ & $P_{\text{DD}}$ & \multicolumn{1}{l|}{1\%} \\
 & \multicolumn{1}{r|}{W} & 0\% & 0\% &  \multicolumn{1}{l|}{100\%} \\ \cline{3-5} 
\end{tabular}
\caption{A conceptual transition matrix for the Markovian defaults technique, wherein the rates $P_{\text{PP}}$ and $P_{\text{DD}}$ are to be systematically varied.} \label{tab:markov_transmatrix}
\end{table}

%% file: 4-Results.tex
\section{Computational results of recovery optimisation}
\label{sec:ch4}

In this section, the LROD-procedure is demonstrated and tested across a wide array of credit risk scenarios generated using the testbed described in \autoref{sec:sim_techs}. The computational results are grouped below by technique, followed by suggestions for applying the LROD-procedure on real-world data.

\subsection{Random defaults}
\label{sec:simulation_study_randomdefaults}

\begin{figure}[H]
\centering
\begin{subfigure}{0.497\textwidth}
    \caption{Using $(4,g_1)$-truncation}
    \centering\includegraphics[width=1\linewidth,height=0.28\textheight]{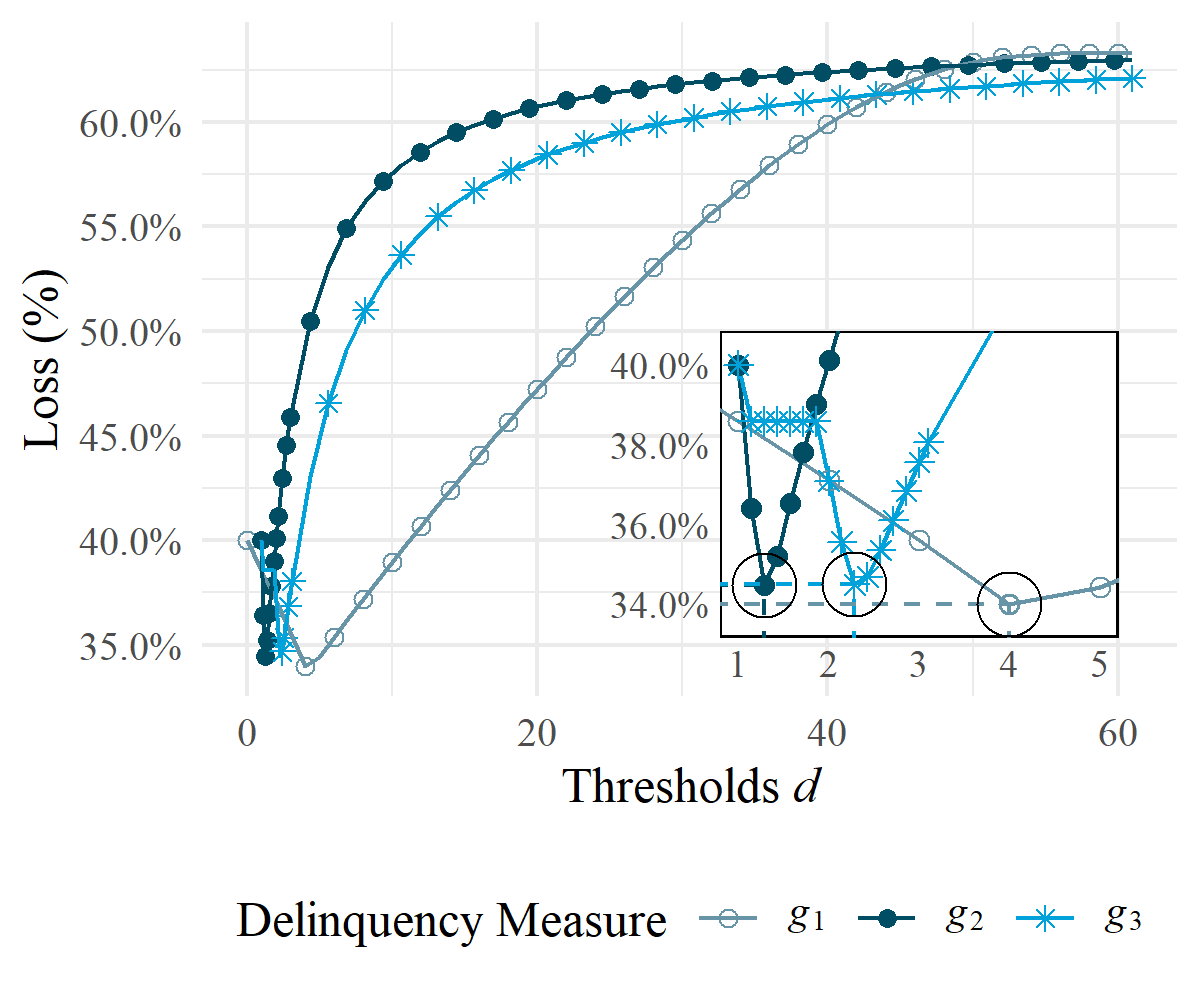}\label{fig:LossThresh_RandomDefaults_a}
\end{subfigure}
\begin{subfigure}{0.497\textwidth}
    \caption{Using $(6,g_3)$-truncation}
    \centering\includegraphics[width=1\linewidth,height=0.28\textheight]{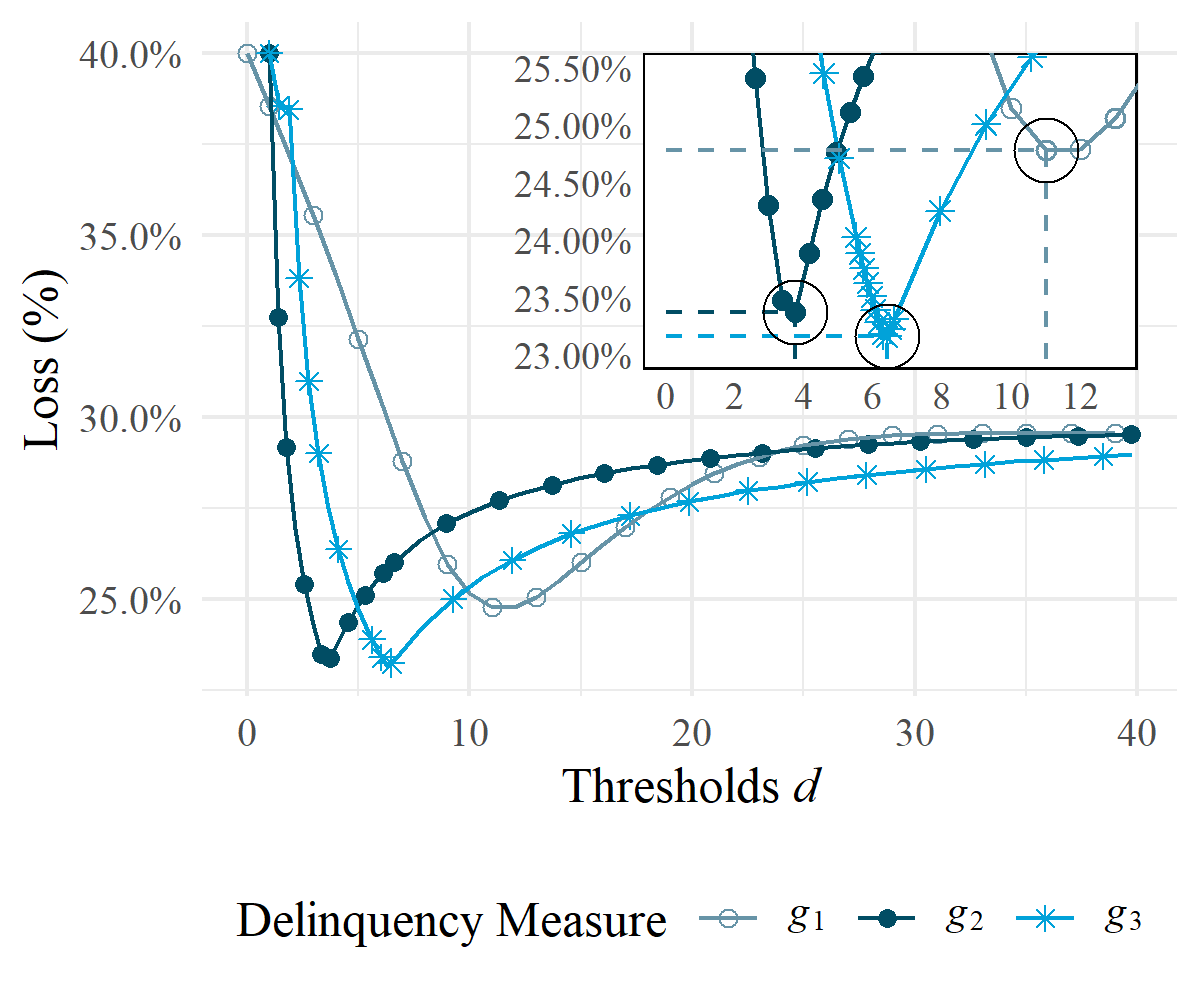}\label{fig:LossThresh_RandomDefaults_b}
\end{subfigure}
\caption{Losses (as a proportion of summed principals) across thresholds $d$ by measure $g \in \big\{ g_1,g_2,g_3 \big\}$ using the random defaults technique. In (a), loans are $(4,g_1)$-truncated, while they are $(6,g_3)$-truncated in (b). In both cases, the zoomed plots show that global minima occur at or near the truncation point, $d=k$.}\label{fig:LossThresh_RandomDefaults}
\end{figure}

This technique leverages $(k,g)$-truncation to control the portfolio generation itself, thereby serving as a sanity check when testing the optimisation results and its underlying logic. Intuitively, the lowest loss should be at threshold $d=k$, since receipts are zeroed after having breached $k$ by design. As an illustration, $(4,g_1)$-truncation is applied in \autoref{fig:LossThresh_RandomDefaults_a}, which shows the lowest loss occurs at $d=4$ for $g_1$ as expected. However, the choice of $g\in \big\{ g_1,g_2,g_3 \big\}$ when truncating introduces bias in the timing of cash flows, such that this $g$ will likely contain the lowest loss as well. This is demonstrated in \autoref{fig:LossThresh_RandomDefaults_b} when using $(6,g_3)$-truncation instead, where the minimum loss now occurs approximately at $d=k=6$ for $g_3$. Whilst seemingly artificial, truncation is merely used as an intuitive testing tool. However, the notion of truncation is plausibly similar to default contagion during a real-world economic downturn, during which borrowers may default systematically at some level of accrued delinquency $k$ on average.

\begin{figure}[H]
\centering\includegraphics[width=0.7\linewidth,height=0.4\textheight]{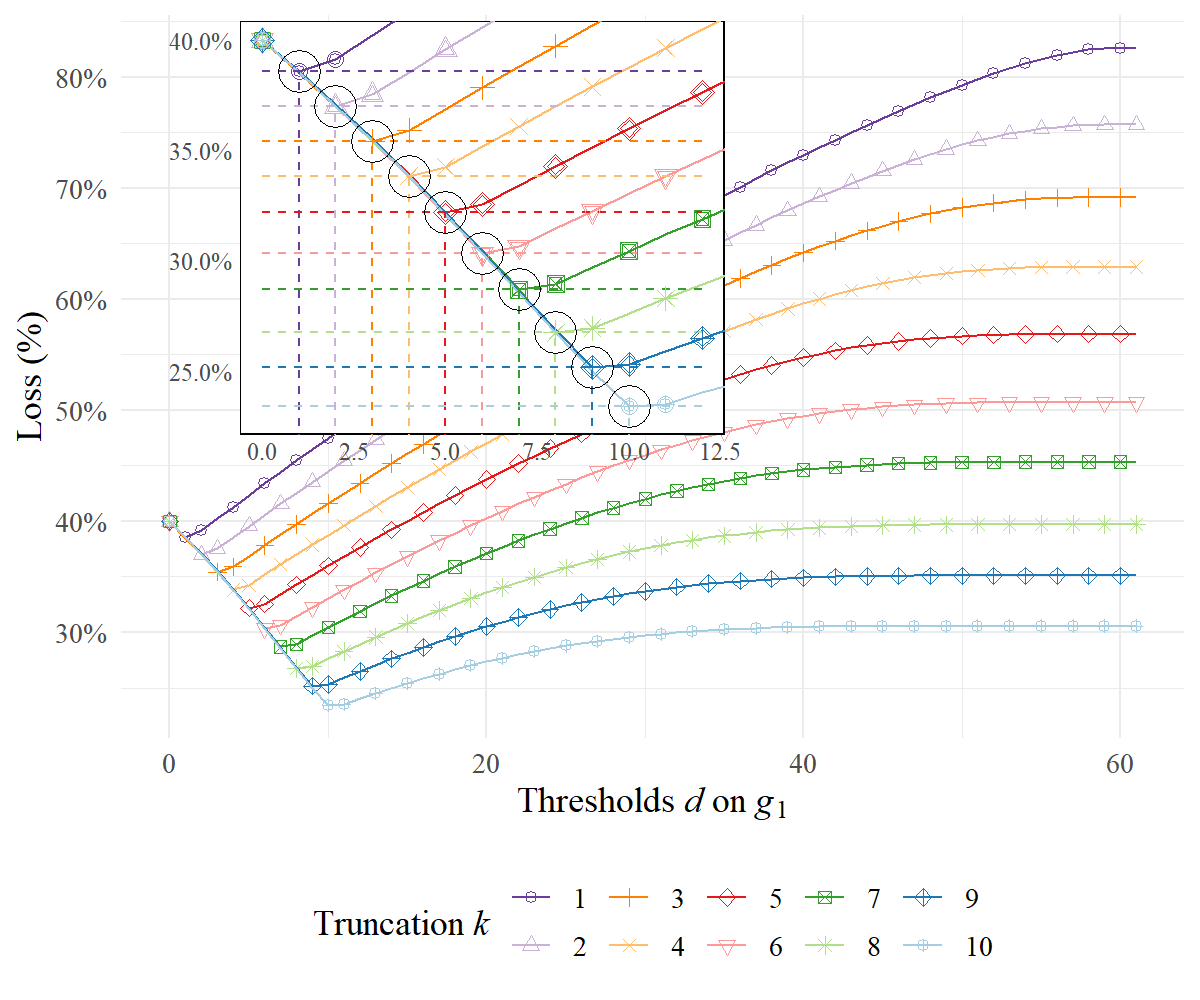}
\caption{Losses across thresholds $d$ for the $g_1$-measure with $(k,g_1)$-truncation, using the random defaults technique. Several truncation points $k=1,\dots,10$ are used, with the zoomed plot confirming that global minima in losses occur at each truncation point $d=k$.}\label{fig:LossThresh_RandomDefaults_ktrunc}
\end{figure}

Minimum losses ought to occur wherever $d=k$ when $(k,g)$-truncating receipts. This intuition is largely confirmed in \autoref{fig:LossThresh_RandomDefaults_ktrunc} wherein truncation parameters $k=1,\dots,10$ are applied during portfolio generation. As a result, loss minima occur consistently at the truncation point $d=k$ as expected, while holding other factors constant. Each increasing value of $k$ also yielded a smaller minimum loss as a result of the overall lessening truncation effect. Since receipts are truncated less frequently as $k$ increases, generated portfolios exhibit overall less delinquency (or credit risk), which explains both lower loss curves and lower loss minima. Although not shown, this result holds similarly for $g_2$ and $g_3$ when used in truncation. Therefore, the optimisation is deemed sensitive to systematic defaults and can react accordingly should the defaulting behaviour of borrowers converge, as simulated by truncation.

\begin{figure}[ht!]
\centering\includegraphics[width=0.7\linewidth,height=0.4\textheight]{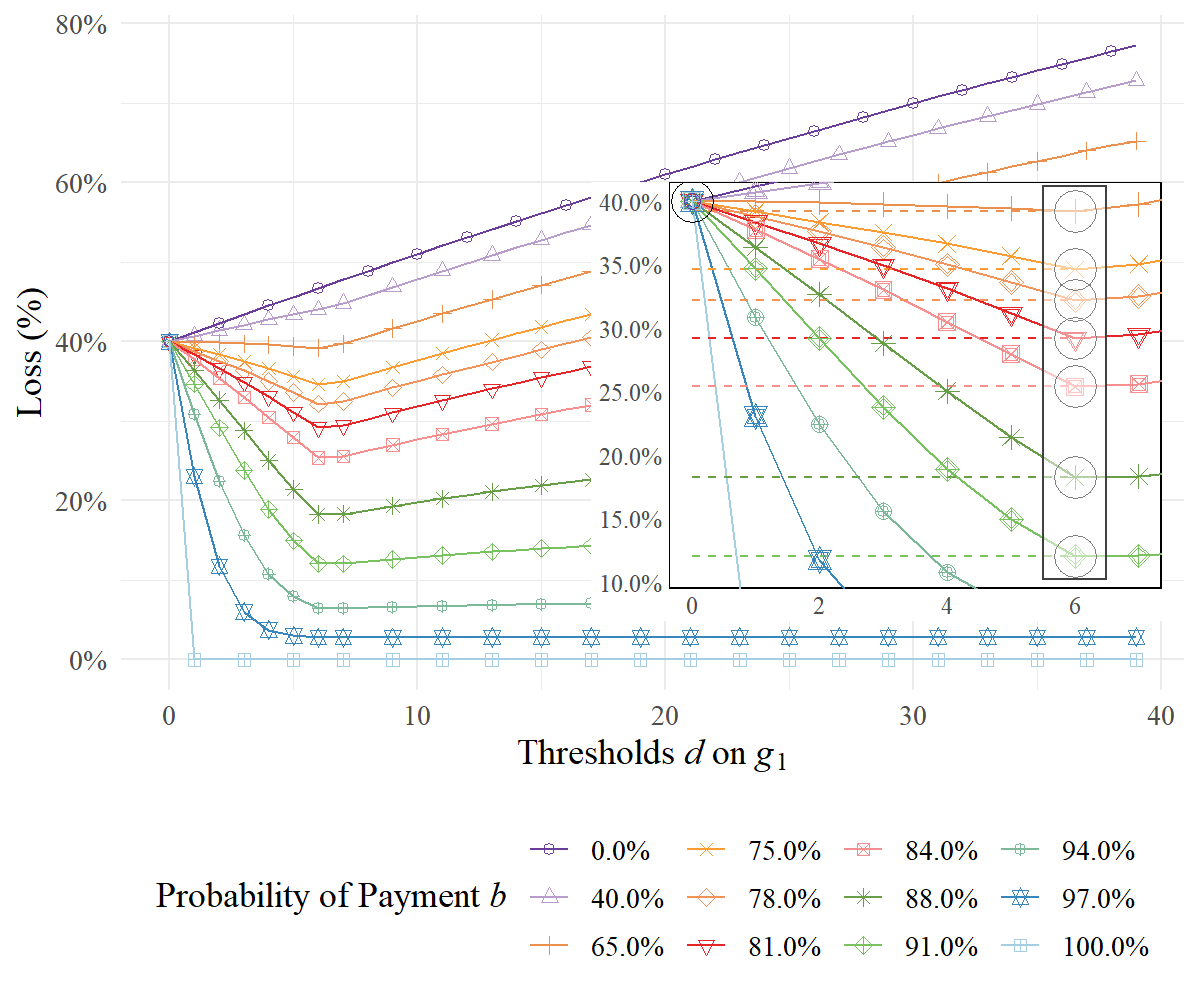}
\caption{Losses across thresholds $d$ for the $g_1$-measure with $(6,g_1)$-truncation, using the random defaults technique and several probabilities of payment $b\in[0,1]$. The zoomed plot shows a smaller range of $0.65\leq b \leq 0.91$ where loss minima occur at the chosen truncation point.}\label{fig:LossThresh_RandomDefaults_probpay}
\end{figure}

Besides truncation, this technique has another parameter that is arguably more relevant: that of the one-period repayment probability $b$. Each value of $b$ corresponds to a particular level of credit risk during portfolio generation. By varying $b$, the effect of credit risk can be broadly tested when optimising loan recovery, as shown in \autoref{fig:LossThresh_RandomDefaults_probpay}. Applying $(6,g_1)$-truncation as a benchmark, loss minima still occur at $d=k=6$ as expected, though only for a certain range of $0.5<b<0.94$. This suggests that optimising loan recovery in practice is infeasible for either very risky loan portfolios or near riskless portfolios. In particular, the two boundary cases of $b=0$ and $b=1$ in \autoref{fig:LossThresh_RandomDefaults_probpay} support this idea in that loans should be forsaken at the outset when $b=0$, as evidenced by the loss minimum at $d=0$, since all receipts will be zero-valued by design. Conversely, if there is no credit risk, i.e., $b=1$, then no loss is made at any $d>0$ and loan recovery itself becomes a moot point. These computational results can directly translate into practical value when estimating the parameter $b$ from a real-world portfolio, as well as estimating the extent of any underlying truncation effect.

\begin{figure}[ht!]
\centering\includegraphics[width=0.7\linewidth,height=0.4\textheight]{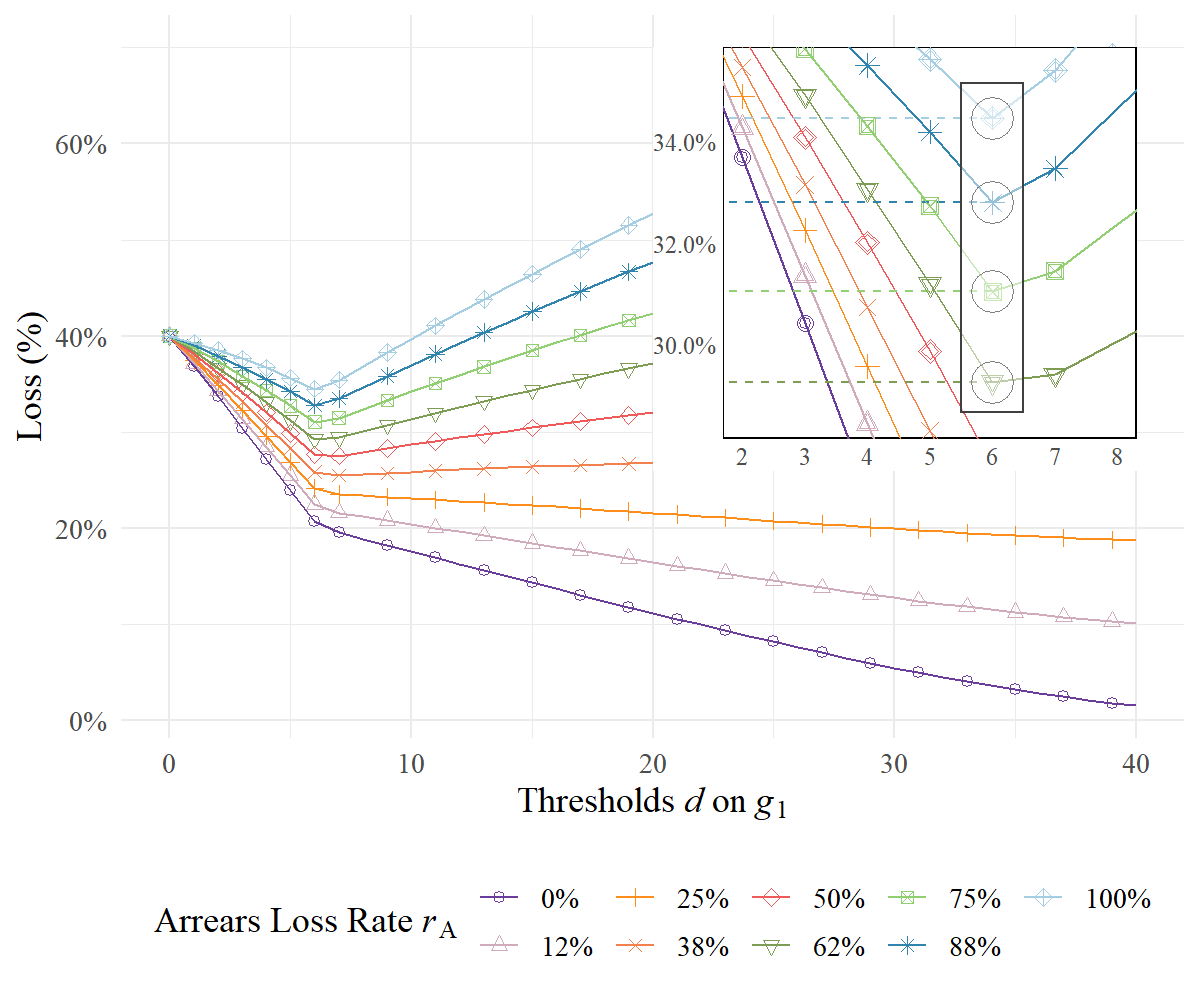}
\caption{Losses across thresholds $d$ for the $g_1$-measure with $(6,g_1)$-truncation, using the random defaults technique and several arrears loss rates $r_A\in[0,1]$. The zoomed plot shows a smaller range of loss rates $0.62 \leq r_A \leq 1$ where loss minima occur at the chosen truncation point.}\label{fig:LossThresh_RandomDefaults_LossRate}
\end{figure}

Intuitively, the loss experience (or LGD) of a particular portfolio ought to affect the results of recovery optimisation as well, especially when considering loan security in the event of default. This is testable by varying the loss rate $r_A$ during portfolio generation while holding other factors constant, as illustrated in \autoref{fig:LossThresh_RandomDefaults_LossRate} using $g_1$ (though similar results hold for $g_2$ and $g_3$). As a proxy for more secure portfolios, smaller values of $r_A$ lead to flatter loss curves, until reaching a point where recovery optimisation becomes infeasible. Conversely, larger values of $r_A$ yield loss curves with a greater `bend' at the chosen truncation point, which signifies the greater risk involved with more unsecured portfolios. Since $b$ is held constant, one can conclude that once default does occur, the viability of recovery optimisation only increases with the risk of loss, which is intuitively sensible. This is to say that unsecured portfolios will likely benefit even more from recovery optimisation than secured portfolios.

\subsection{Markovian defaults}
\label{sec:simulation_study_markoviandefaults}

\begin{figure}[ht!]
\centering\includegraphics[width=1\linewidth,height=0.28\textheight]{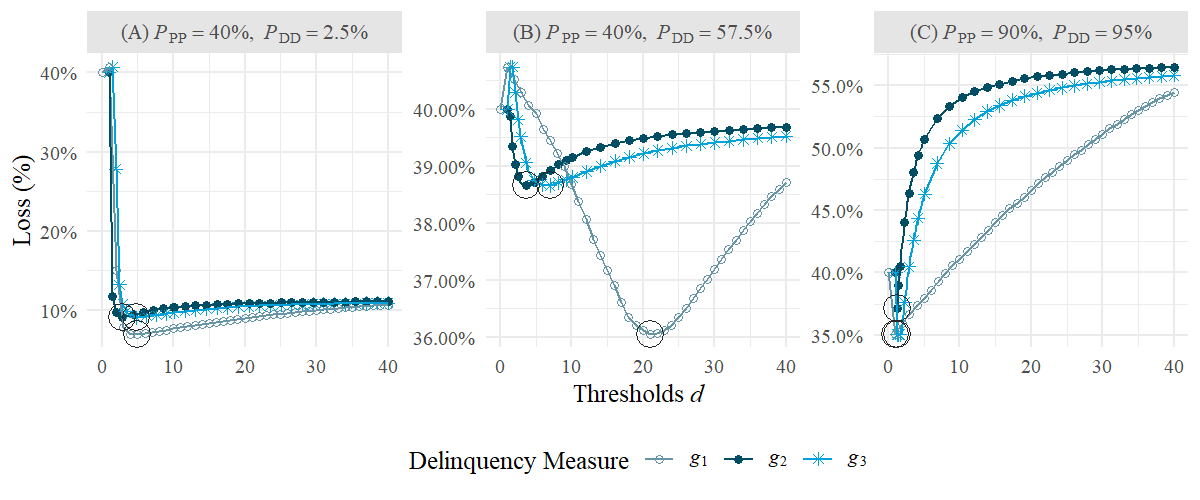}
\caption{Losses across thresholds $d$ by measure $g \in \big\{ g_1,g_2,g_3 \big\}$ using the Markovian defaults technique to generate different loan portfolios. Each panel explores a specific setting of the transition matrix, using the titular probabilities within the matrix defined in \autoref{tab:markov_transmatrix}. Encircled points indicate loss minima at associated thresholds $d^{(g)}$.}\label{fig:LossThresh_Markovian_Experiments}
\end{figure}

This technique affords greater flexibility in generating portfolios with more sporadic repayment histories. Accordingly, the LROD-procedure is demonstrated in \autoref{fig:LossThresh_Markovian_Experiments} using some of the parametrisations of the underlying Markov chain that yield optima across all delinquency measures. Evidently, the $g_1$-measure appears to outperform the other measures since it yields the lowest loss within each of these settings, including a number of other parametrisations not shown. However, summarily concluding the supremacy of $g_1$ across \textit{all} portfolios would be disingenuous. It is still possible that some real-world portfolios may be better served using measures other than $g_1$ within the LROD-procedure (or more broadly in risk management). In addition, the current objective is \textit{not} to determine the best measure conclusively. Indeed, conducting such an empirical study would require expansive real-world data on all types of portfolios across the risk spectrum, which is prohibitively impractical at this stage. That said, the $g_1$-measure is henceforth used in this section given its supremacy in this instance.

\begin{figure}[ht!]
\centering\includegraphics[width=1\linewidth,height=0.64\textheight]{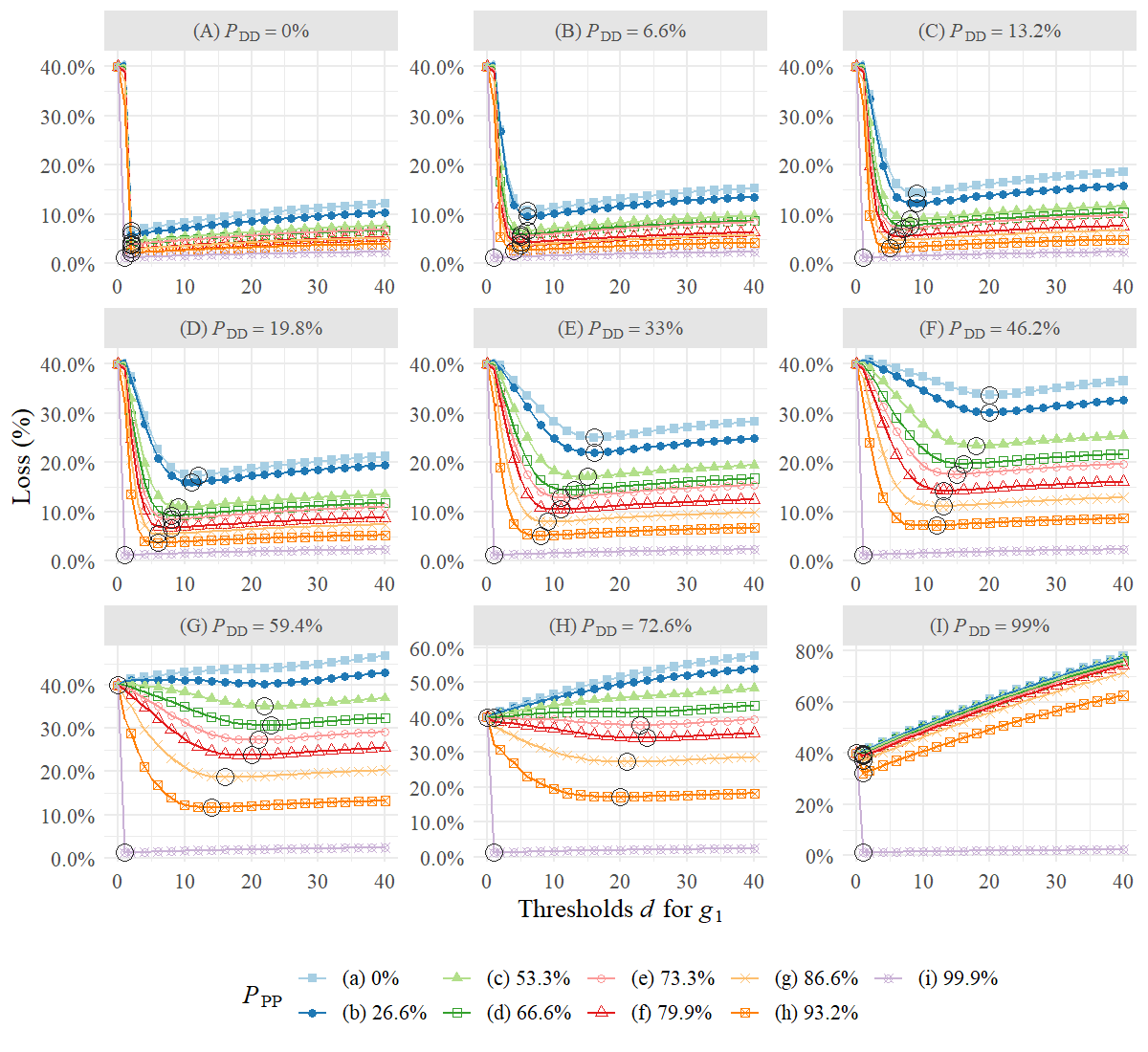}
\caption{Losses across thresholds $d$ for the $g_1$-measure using the Markovian defaults technique with several transition rates $P_{\text{PP}}\in[0,1]$ and $P_{\text{DD}}\in[0,1]$. Encircled points indicate loss minima at associated thresholds $d^{(g_1)}$.}\label{fig:LossThresh_Markovian}
\end{figure}

Using this technique, we devise a broad iterative scheme to generate portfolios systematically across the entire credit risk spectrum, as measured with $g_1$. In particular, $P_{\text{DD}}$ is held constant at a certain value while varying $P_{\text{PP}}$, followed by fixing $P_{\text{DD}}$ to a different value and varying $P_{\text{PP}}$ again, and so on. This scheme allows for suitably varying the transition matrix in \autoref{tab:markov_transmatrix} using fixed intervals, with some of the resulting loss curves and associated loss minima presented in \autoref{fig:LossThresh_Markovian}. The subplots in both panels (A) and (I) represent boundary cases that confirm intuition. Specifically, panel (A) demonstrates recovery optimisation for portfolios with highly transitive delinquency states such that accounts immediately exit this state in the next period, once entered. Accordingly, the loss curves increasingly resemble a near risk-less case as the value of $P_{\text{PP}}$ tends towards 1, which is similar to setting $b=1$ in \autoref{fig:LossThresh_RandomDefaults_probpay} when using random defaults. In turn, recovery optimisation itself becomes progressively infeasible in tandem with $P_{\text{PP}}$ approaching 1. Conversely, panel (I) showcases the loss curves of extremely risky portfolios, which are again similar to setting $b=0$ in \autoref{fig:LossThresh_RandomDefaults_probpay} as $P_{\text{PP}}$ approaches 0. More importantly, the fact that loss minima occur at very small thresholds agrees intuitively with cutting losses sooner rather than later, especially for extreme default risk.

The remaining panels in \autoref{fig:LossThresh_Markovian} are perhaps the most instructive. As the delinquency state becomes more absorbing (or less transient), i.e., moving from panel (B) to (F), the loss-optimal thresholds $d^{(g_1)}$ become increasingly staggered across both axes. This is to say that $d^{(g_1)}$ becomes progressively more sensitive to both the threshold $d$ and the value of $P_{\text{PP}}$. Moreover, it is sensible that ever greater losses (at $d^{(g_1)}$) are associated with lower values of $P_{\text{PP}}$ since the latter implies less time being spent in the paying state, even as the delinquency state becomes less transient. Furthermore, consider that $d^{(g_1)}$ increases in threshold-value when $P_{\text{PP}}$ decreases and $P_{\text{DD}}$ increases, i.e., moving from curve (i) down to curve (a) whilst moving across panels (B) to (F). This suggests that gradually postponing loan recovery is the better strategy even as delinquency becomes more likely, at least up until a certain point, in this case, panel (G). However, this suggestion is counter-intuitive since one would rather cut losses sooner than later when risk supposedly increases, which implies selecting lower thresholds instead. 
Two factors help explain this phenomenon. Firstly, the relevant portfolios are increasingly turbulent by design when $P_{\text{PP}}$ changes from higher to lower values in each successive panel. The effect hereof is that loans start to oscillate quite rapidly between the paying and delinquent states as $P_{\text{PP}}$ decreases. The slightly increased rate of absorption into the delinquent state (when moving across panels) is not sufficient to support earlier loan recovery as intuition would otherwise suggest, especially so when an account still frequently exits the delinquent state. This has the side-effect of muting the severity of `default', which is plausible when curing from `default' itself becomes increasingly likely due to the same turbulence. Therefore, the associated opportunity cost of forsaking the loan earlier is too high when future repayments are still likely to be received over the longer run, albeit sporadic. Accordingly, greater turbulence in a portfolio requires greater patience to collect upon these repayments, which is why postponing loan recovery (by virtue of $d^{(g_1)}$ increasing) would be loss-optimal. Secondly, even if $d^{(g_1)}$ increases in value, the associated loss minimum reassuringly increases alongside $P_{\text{DD}}$, as expected from more turbulent and riskier portfolios. 

There is little need for applying $(k,g)$-truncation on these results since the Markovian technique already has a realistically-set write-off state that achieves the same effect. While additional truncation will surely confound the results, $(12,g_1)$-truncation is experimentally applied in the interest of completeness. The results (not shown) are largely similar to that of random defaults in that loss minima still occur at or near $k=12$ across most portfolios. The exceptions are the two boundary cases, i.e., at or close to panels (A) and (I). Furthermore, the Markovian technique is especially geared towards generating "regime-switching" portfolios where accounts suffer from episodes of delinquency that vary in length, as controlled by the state probabilities. In this regard, episodic delinquency is more common a phenomenon in practice than one would think, which is why investigating recovery optimisation for these cases is more valuable than exploring explicit truncation/write-off any further in this section.

\subsection{Applying the LROD-procedure on real-world data}
\label{sec:simulation_study_application}

The steps in \autoref{sec:loss_proc} require data to be in a longitudinal-format, having measured delinquency in retrospect across all accounts and time (usually monthly), based on expected instalments and actual receipts. Letting the contractual term, loan and risk-free rates, and even the loss rates vary across the portfolio ought not to impede the practical use of the LROD-procedure. However, the portfolio is assumed to be fully observed (or 'completed') in this study, with little consideration given to any right-censoring and its effect on the receipt history of an account. This particular avenue is further explored in \citet{botha2020paper2}, thereby demonstrating the empirical use of the procedure on real-world data. That said, simply excluding incomplete accounts from the dataset can sidestep this possible issue, though at the cost of a reduced sample size. The effect hereof will likely vary based on the typical tenure of the loan product.

The results, particularly those from \autoref{sec:simulation_study_markoviandefaults}, can easily translate into practical value with relatively little analytical effort. For example, one can fit the same three-state Markov chain on a real portfolio's delinquency progressions, just to obtain the associated transition rate estimates. In turn, these estimates can be used as a rough guide in finding a corresponding loss curve amongst all those presented in \autoref{fig:LossThresh_Markovian}, i.e., a look-up exercise. The associated optimised threshold can provide a high-level idea of recovery optimisation, provided the assumptions are reasonably met. That said, applying the LROD-procedure remains the imperative in order to capture all idiosyncrasies of a particular portfolio and the prevailing market conditions.

%% file: 5-Conclusion.tex
\section{Conclusion}
\label{sec:ch5}

We explore a more fundamental meaning of loan `default' by only using $d$ as a variable threshold upon the domain of a delinquency measure $g$. Though different from current practices, this reinterpretation of `default' better aligns with the rather probabilistic idea of breaching a certain "point of no return", having exceeded $d$ on $g$. In principle, keeping the loan any longer beyond this point becomes sub-optimal to abandoning it and recovering the maximum instead. To this end, we contribute a novel optimisation procedure as an expert system to help find the \textit{ideal} time for debt recovery during loan life, based on accrued delinquency. This so-called LROD-procedure weighs two competing interests against each other: the prospect of reaping future revenue from troubled loans versus the cost of retaining these loans any further. In principle, each $(g,d)$-configuration serves as a candidate collection policy that has a "net cost" if applied to a portfolio. The overall portfolio loss is then iteratively calculated across all such policies using the procedure's inner loss model ($L_g$). Doing so forms a loss curve for each $g$ that can be inspected for an optimal threshold at which the lowest loss occurs, thereby concluding the optimisation. In addition, the LROD-procedure is formulated in such a way that it can be used with multiple loan delinquency measures. This facilitates the objective testing of alternative measures, e.g., those provided in the appendix, that may better suit the recovery optimisation (or even broader risk management) of a portfolio. That said, the study objective is not to establish the best measure conclusively, which would likely be a data-intensive and costly endeavour.

Regarding results, a simple simulation-based setup is first described in which the LROD-procedure (and its goal of recovery optimisation) is closely examined from "first principles". Using this setup as a testbed, a broad computational study is conducted wherein basic amortising loan portfolios are systematically generated by varying the simulation parameters, though still constrained by expert judgement. Having spanned the entire credit risk spectrum, the computational results show that optimising the recovery decision's timing is viable across most risk levels, except at the extremes. We further demonstrate that optimised recovery times are sensitive to systematic defaults that may structurally affect a portfolio during an economic downturn. Another factor is that of collateral and the portfolio's loss experience (or LGD), insofar that optima were successfully found across most of the loss spectrum. Moreover, recovery optimisation seems to become an increasingly viable practice as the risk of loss increases.

In addition, recovery optimisation is tested on more turbulent portfolios wherein borrowers repay intermittently, thereby causing episodic delinquency. Once accounts oscillate rapidly between paying and nonpayment, `default' itself diminishes in severity, especially when curing also becomes more likely as a result of the very same turbulence. Accordingly, we found that optimised thresholds increased in value as turbulence develops, though only up to a point. Postponing loan recovery in tandem with greater turbulence is therefore strategically optimal since it allows greater scope to collect upon these repayments, albeit sporadic. As a secondary contribution, the testbed itself can serve as a valuable tool in exploring the strategic viability of the LROD-procedure. Once appropriately parametrised, the testbed can generate a wide variety of portfolios, which allows a bank to investigate (at least preliminarily) the prospects of recovery optimisation for a certain type of portfolio. Ultimately, the LROD-procedure can be used to tweak existing collection policies and, perhaps in time, default definitions themselves.

Future studies can focus on refining the LROD-procedure using real-world portfolio data. So-called `incomplete' portfolios, i.e., those wherein many loans have not yet reached contractual maturity, may prove a challenge for recovery optimisation at this stage. The simplest solution would be to exclude the incomplete accounts, though unfortunately reducing the sample size as well. Alternatively, one can perhaps explore an appropriate forecasting approach in future work. Furthermore, homogeneity is currently assumed in that the optimised threshold is a portfolio-wide criterion. However, exploring segmentation schemes may be worthwhile such that the LROD-procedure yields an ideal threshold for each identified segment within the portfolio. Lastly, the current loss model $L_g$ can be refined by incorporating historical loss experiences and transforming it into a more dynamic component. As an example, calculating the realised LGD generally requires a specific point of entering `default'. From this point, cash flows are observed during its workout up to the applicable write-off point. By introducing $d$ as the $(g,d)$-default state, the starting points of cash flows will naturally vary with $d$, thereby impacting the LGD calculation itself for each $(g,d)$-policy. Intuitively, longer or shorter workout periods will affect the loss experience, which will influence recovery optimisation based on the study results. This particular refinement will likely intersect with the existing literature on credit loss modelling and IFRS 9, which as a field is currently quite in vogue.

%% file: 6-Appendix.tex
\section{Appendix}
\label{app:delinquency_measures}

Three mathematical quantities are presented as delinquency measures in this appendix. Firstly, a variant of the widely-used number of payments/months in arrears, called the $g_1$-measure (or \textit{CD}-measure), is refined into a more robust measure in \autoref{sec:cd_measure} using a weighting scheme. Secondly, a more concise algorithm is contributed in \autoref{sec:md_measure} that creates the Macaulay Duration Index from \citet{sah2015}, called the $g_2$-measure (or \textit{MD}-measure), which is an index of the weighted average time to recover the capital portion of a loan. Lastly, a modified version of $g_2$ of our own invention is introduced in \autoref{sec:dod_measure}, called the $g_3$-measure (or \textit{DoD}-measure), which incorporates the sizes of disrupted cash flows into delinquency assessment.

\subsection{Contractual Delinquency (\textit{CD}): the \texorpdfstring{$g_1$}{Lg}-measure}
\label{sec:cd_measure}

Days past due (DPD) from accountancy practices is commonly used in constructing a delinquency measure $g$, whereby the unpaid portion of a loan's instalment is binned into increasingly severe groups as each 30-day calendar month lapses: 30 days, 60 days, 90 days, and so forth, as discussed in \citet{cyert1962}. More formally, $g$ is defined as the function $g_0(t)=f(A_t/I)$ where $A_t$ denotes the accumulated arrears amount at discrete time $t$, $I$ is the fixed instalment, and $f$ is a chosen rounding function that maps the given input to the number of payments in arrears as the output. A common choice of $f$ is the ceiling function, whereby the input is simply rounded upwards to the nearest integer. 

However, this rounding scheme is quite stringent in that even a small difference $I_t-R_t=\epsilon < \text{ZAR} \, \, 1.00$ will increase the delinquency measurement, purely due to rounding. Depending on the volatility in $R_t$ over time, it is punitive to penalise a borrower when $\epsilon$ is but a few cents. That said, a sensible boundary on $\epsilon$ must be applied, otherwise the idea of delinquency becomes meaningless. Should $A_t/I$ simply be rounded to the nearest integer instead, then a change in $g_0(t)$ over time $[t_1,t_2]$ depends entirely on whether $A_t/I$ is above or below 50\%. This implied `threshold' seems arbitrary, inflexible, and certainly at odds with the risk-based practices of a bank. Lastly, constructing $g_0$ in practice quickly becomes cumbersome when the instalment is linked to an interest rate that varies over time, which is common for secured lending.

There are two additional pitfalls to the $g_0$-measure. Firstly, overall measurement can become lagged by one (or more) periods when a significant overpayment is immediately followed by a severe underpayment the following month, purely due to the chosen $f$. Secondly, if $A_t$ accumulates interest on itself or attracts any fees, then $g_0$ can become `corrupted' due to its inherent reliance on $A_t$. The potential exists for $g_0$ to change in value, not due to a fundamental breakdown in trust, but as a result of the lender's own pricing structure or system constraints, which may artificially inflate the $g_0$-value. Moreover, the rounding scheme itself may exacerbate this effect. In both of these cases, the apparent "measurement error" in $g_0$ can adversely affect the true accuracy of models predicting default risk.

Therefore, a more comprehensive variant, called the \textit{CD}-measure, is presented here that circumvents these challenges. Let the receipt vector be $\boldsymbol{R}=[R_0,R_1,\dots,R_T]$ with its elements (or receipts) $R_t\geq 0$, and let the instalment vector be $\boldsymbol{I}=[I_0,I_1,\dots,I_T]$ with its elements $I_t>0$. Both vectors are defined for a specific loan account across its discrete time periods $t=0,\dots,T$, with $t=0$ representing the origination point and $T$ denoting the tenure (or current loan age). Note that $T$ may exceed the contractual term $t_c$, especially in cases of extreme delinquency. The repayment ratio $h_t\in[0,\infty)$ is then defined as \begin{equation} \label{eq:repay_ratio}
h_t = \frac{R_t}{I_t} \quad \forall \ t=1,\dots,T \quad \text{and} \quad h_0=0 \ .
\end{equation} One can specify a certain threshold $z \in [0,1]$ for $h_t$, above which an account at time $t$ is considered current and beneath which it is considered delinquent. Note that $z=90\%$ is assumed in this study purely as an illustration, though the lender should certainly adjust this $z$ accordingly. Next, a Boolean-valued decision function $\mathcalz{d}_1(t) \, \in \{0,1\}$ is defined for $t=1,\dots,T$, using Iverson brackets $[a]$ that outputs 1 if the enclosed statement $a$ is true, and 0 if false, as \begin{equation} \label{eq:decide_1}
\mathcalz{d}_1(t)=\big[h_t<z \big] \ .
\end{equation}

Memory of past delinquency is introduced by defining another integer-valued function $m(t) \in \{-1,0,1,\dots\}$ for $t=1,\dots,T$, which outputs the reduction in accrued delinquency (if any), as \begin{align} \label{eq:del_reduction}
m(t) & =\left( \floor*{\frac{h_t}{z}} - 1 \right) \Big(1-\mathcalz{d}_1(t) \Big) - \mathcalz{d}_1(t) \nonumber \\
	 & = \floor*{\frac{h_t}{z}} \Big(1 - \mathcalz{d}_1(t) \Big) - 1  \ .
\end{align} This function $m(t)$ gives the magnitude by which the measured delinquency at time $t$ should be reduced (if at all) in catering for past delinquency. When overpaying, i.e., $R_t>I_t$, the ratio between $h_t$ and $z$ in \autoref{eq:del_reduction} signifies the total number of `payments' by which accrued delinquency should be decreased, as weighed by $z$. The rounding problem from $g_0$ is resolved in this measure when dividing by $z$ since its specified value reflects the lender's tolerance towards underpayment by design. Accordingly, taking the floor of $h_t / z$ does not detract and merely enforces an integer-valued scale in the eventual measure. Furthermore, the currently-owed instalment should be recognised first before reducing any accrued delinquency, which is achieved by subtracting one instalment. For sufficient underpayment, i.e., $R_t < zI_t$, the delinquency is sensibly increased by one payment, which resolves to $m(t)=-1$ when $\mathcalz{d}_1(t)=1$.

To indicate previous cases of delinquency using $g_1$ at time $t-1$, let $\mathcalz{d}_2(t) \, \in \{0,1\}$ be another Boolean-valued decision function for $t=1,\dots,T$, which is defined using Iverson brackets again, as \begin{equation} \label{eq:decide_2}
\mathcalz{d}_2(t)=\big[g_1(t-1) = 0 \big] \ .
\end{equation} The reduction in delinquency $m(t)$ at time $t$ is subtracted from delinquency as measured at the previous period $t-1$, thereby giving net delinquency. The integer-valued \textit{CD}-measure $g_1(t) \geq 0$ for $t=1,\dots,T$ is then recursively expressed as \begin{equation} \label{eq:g_1}
g_1(t)= \max{ \Bigg[0, \quad \mathcalz{d}_1(t)\mathcalz{d}_2(t) \ + \ \big(1 - \mathcalz{d}_2(t) \big) \bigg(g_1(t-1) - m(t) \bigg) \Bigg]} \ .
\end{equation} Note the necessary starting condition of $g_1(0)=0$, since a newly-disbursed loan cannot yet be delinquent. The output for $g_1$ is best interpreted as the $z$-\textit{weighted} number of payments in arrears, weighed by the lender's tolerance (or appetite) towards accrued arrears. Since delinquency only increases if $h_t<z$ by definition, a higher value of $z$ effectively translates to greater risk-aversion, and \textit{vice versa} for lower $z$-values.

\subsection{Macaulay Duration (\textit{MD}): the \texorpdfstring{$g_2$}{Lg}-measure}
\label{sec:md_measure}

The Macaulay Duration Index, recently introduced in \citet{sah2015}, is based on bond duration, i.e., the weighted average time to recover the capital portion of a loan. This measure incorporates the loan's interest rate as well as the arrears balance weighted by the time value of money. It is constructed as the ratio between the actual and expected loan duration, reworked as the $g_2$-measure in this study. However, the values of $g_2$ are incomparable to those of $g_1$ since both the domains and meanings differ.

Let $\Delta{_t}=I_t - R_t$ be the difference between the instalment $I_t$ and the receipt $R_t$ at every time point $t=0,\dots,T$ during the life of a loan, including at disbursement $t=0$ to capture any applicable initiation fees. Considering the time value of money, let $v_j=(1+r)^{\, -j}$ be a discounting function that uses a nominal monthly interest rate $r$. In addition, let $\delta$ be the continuously compounded rate with its nominal variant $\delta{^{(p)}}=\delta/p$ and with an annual compounding period $p=12$. Let $L_P$ denote the loan amount (or principal) that is to be amortised. Ordinarily, the Macaulay Duration is calculated (perhaps once) at origination as the weighted average time to recover sunk capital from future cash flows. However, here it is recursively calculated instead at each subsequent period $t=0,\dots,T$ across the remaining $m$ instalments as at each $t$. Naturally, this \textit{expected duration} quantity, denoted as $\mathcalz{f}_{\, ED}$, tends towards zero over time as it nears the end of loan life, expressed as \begin{equation} \label{eq:exp_dur}
\mathcalz{f}_{\, ED}(t) = \sum_{m=t}^{T}{ \left[ \left( \frac{I_m v_{m-t}}{L_P} \right) \left( \frac{m-t}{p} \right) \right] } \quad \forall \ t=0,\dots,T \ .
\end{equation} 

However, \autoref{eq:exp_dur} assumes that instalments $\boldsymbol{I}$ are free of uncertainty. When substituting these instalments with the actual receipts $\boldsymbol{R}$, a significant difference is intuitively expected. Moreover, it becomes necessary to track the arrears balance as it develops (if it does) over the loan life. In line with \citet{sah2015}, any arrears at any time are added to the last expected (contractual) instalment at $t=t_c$, since it represents the last contractual opportunity to repay any such arrears, short of the lender intervening and restructuring the loan. This last instalment is then recursively updated for each subsequent period $t$, denoted by the vector $\boldsymbol{I}'$, which equals instalments $\boldsymbol{I}$ at first. Lastly, the \textit{actual duration} $\mathcalz{f}_{\, AD}(t)$ is also recursively calculated for each subsequent period $t$. This is illustrated using pseudo-code in \autoref{alg:md_measure}.

\begin{algorithm}
  \caption{Calculating $g_2$}\label{alg:md_measure}
   \begin{algorithmic}[1]
     \State $\boldsymbol{I}':=\boldsymbol{I}$, where $\boldsymbol{I}=\big[I_0,\dots,I_T \big]$  and $T \leq t_c$
     \State $\mathcalz{f}_{\, AD}(0):=\mathcalz{f}_{\, ED}(0)$
     \For{$t=0,\dots,T$} \Comment{such that $T\leq t_c$}
     	\State $I_{(T)}' := I_{(T)}' \ + \ \Delta{_t} \left(1 + \frac{\delta{^{(p)}}}{p} \right)^{T-t}, \quad \forall \ t=1,\dots,T $
     	\Comment{Add any arrears to $I_{(T)}'$}
        \State $\mathcalz{f}_{\, AD}(t):= \sum_{m=t}^{T \ | \ T\leq t_c}{ \left[ \left( \frac{I_m' v_{(m-t)} }{L_P} \right) \left(\frac{m-t}{p} \right) \right]}, \quad \forall \ t=1,\dots,T $
     \EndFor
   \end{algorithmic}
\end{algorithm}

Finally, the real-valued Macaulay Duration (\textit{MD}) measure $g_2(t)\geq 0$ is then defined as the ratio between the actual duration and the expected duration at time points $t=0,\dots,T-1$, which is expressed as \begin{equation} \label{eq:g_2}
g_2(t) = \frac{\mathcalz{f}_{\, AD}(t)}{\mathcalz{f}_{\, ED}(t)} \ .
\end{equation}

\subsection{Degree of Delinquency (\textit{DoD}): the \texorpdfstring{$g_3$}{Lg}-measure}
\label{sec:dod_measure}

From a cash flow perspective, an ideal delinquency measurement should penalise the non-payment of a larger loan's instalment to a greater degree than that of a smaller loan's instalment, given the relatively larger impact on a bank's cash flow. Furthermore, the differences in risk concentration between a larger number of small loans versus a small number of larger loans should also be incorporated by the ideal delinquency measure. As a possible solution, the actual duration $\mathcalz{f}_{\, AD}(t)$ from \autoref{eq:g_2} can be altered such that the eventual $g_2(t)$ is greater for larger loans than for smaller loans by defining an appropriate multiplier.

Note that $g_2$ is only defined up to the contractual term $t_c$. However, delinquency can continue even past its contractual term $T\geq t_c$, likely due to persisting underpayment. Ignoring loan write-off policies for the moment, let $\mathcalz{d}_3(t) \, \in \{0,1\}$ be a Boolean-valued decision function that returns 1 if the given time point $t$ precedes the contractual term $t_c$, and 0 if otherwise. Using Iverson brackets, this is expressed as \begin{equation} \label{eq:decide_3}
\mathcalz{d}_3(t)=\big[t \leq t_c \big] \ .
\end{equation} When $t > t_c$, any arrears can clearly no longer be added to the last contractual instalment (since it has lapsed), as was added for $I_{(T)}'$ at $T=t$ when calculating $g_2$ in \autoref{alg:md_measure}. Instead, at least one more payment, albeit out-of-contract, can reasonably be expected at every subsequent period $t : t \geq t_c$ as long as collection efforts are actively pursued. Therefore, delinquency can now be computed up to time $\mathcal{T}$ instead of the previous $T$, with $\mathcal{T}$ either representing the contractual term $t_c$ when $t<t_c$, or becoming a moving target $\mathcal{T} =t$ when $t\geq t_c$. Note that both $\boldsymbol{I}$ and $\boldsymbol{R}$ will incrementally expand with additional elements for as long as collection efforts continue past $t_c$. A revised algorithm is given in \autoref{alg:dod_measure}.

\begin{algorithm}
  \caption{Calculating $g_3$}\label{alg:dod_measure}
   \begin{algorithmic}[1]
     \State $\boldsymbol{I}':=\boldsymbol{I}$, where $\boldsymbol{I}=\big[I_0,\dots,I_T \big]$ and $0<t_c \leq T$
     \State $\mathcal{T}:=t_c$
     \For{$t=0,\dots,T$}
     	\State $\alpha := I_{(\mathcal{T})}'$
        \Comment{This refers to the element at the $\mathcal{T}^{\text{th}}$ position of $\boldsymbol{I}'$}
        \State $\mathcal{T} := t_c\mathcalz{d}_3(t) \ + \ t\bigg(1 - \mathcalz{d}_3(t) \bigg)$
        \Comment{$\mathcal{T}$ is either equal to $t_c$ or to $t\geq t_c$}
        \State $I_{(\mathcal{T})}' := I_{(\mathcal{T})}'\mathcalz{d}_3(t) \ + \ \Delta{_t} \left(1 + \frac{\delta{^{(p)}}}{p} \right)^{\mathcal{T} - t} \ + \ \alpha \bigg(1 - \mathcalz{d}_3(t) \bigg) \left(1 + \frac{\delta{^{(p)}}}{p} \right), \quad \forall \ t=1,\dots,T $
        \State $\beta(m) := m-t+1-\mathcalz{d}_3(t), \quad \forall \ t=1,\dots,T $
        \Comment{Discounting period, see next 3 lines}
        \State $\mathcalz{f}_{\, ED}(t):= \sum_{m=t}^{\mathcal{T}}{ \left[ \left( \frac{I_m v_{\beta(m)} }{L_P} \right) \left(\frac{\beta(m)}{p} \right) \right]}, \quad \forall \ t=0,\dots,T$
        \State $\mathcalz{f}_{\, AD}(t):=\mathcalz{f}_{\, ED}(t)$, $\quad$ for $t=0$
        \State $\mathcalz{f}_{\, AD}(t):= \sum_{m=t}^{\mathcal{T}}{  \left[ \left( \frac{I_m' v_{\beta(m)} }{L_P} \right) \left(\frac{\beta(m)}{p} \right) \right]}, \quad \forall \ t=1,\dots,T $
     \EndFor
   \end{algorithmic}
\end{algorithm}

Afterwards, let $\lambda(L_M,L_P,s)$ denote a multiplier function that inflates $\mathcalz{f}_{\, AD}(t)$ at the period $t$. Let $L_M$ denote the maximum loan size and let $s \in [0,1]$ be a real-valued sensitivity that represents the `strength' at which to apply this inflationary effect. Let $\mathcalz{d}_4(t) \, \in \{0,1\}$ be another Boolean-valued decision function that returns 1 if there is currently any accrued delinquency at $t$, and 0 otherwise, defined using Iverson brackets as \begin{equation} \label{eq:decide_4}
\mathcalz{d}_4(t)=\big[\mathcalz{f}_{\, AD}(t) > \mathcalz{f}_{\, ED}(t) \big] \ .
\end{equation} As a simple example, this multiplier is defined as \begin{equation} \label{eq:lambda_multi}
\lambda(L_M,L_P,s) = s \left(1 - \frac{L_M - L_P}{L_M} \right)  \ .
\end{equation} The inflated variant of $\mathcalz{f}_{\, AD}(t)$, denoted as $\tilde{\mathcalz{f}}_{\, AD}(t)$, is given by \begin{equation} \label{eq:inflated_actdur}
\tilde{\mathcalz{f}}_{\, AD}(t) = \mathcalz{f}_{\, AD}(t) \big( \, \mathcalz{d}_4(t)\lambda(L_M,L_P,s) \ + \ 1  \big)  \ .
\end{equation}

By including $\mathcalz{d}_4(t)$ into $\tilde{\mathcalz{f}}_{\, AD}(t)$ in \autoref{eq:inflated_actdur}, accrued delinquency will not be inflated when overpaying at some period $t$. Finally, the real-valued Degree of Delinquency (\textit{DoD}) measure $g_3(t)\geq 0$ is defined for $t=0,\dots,T-1$ and expressed as \begin{equation} \label{eq:g_3}
g_3(t) = \frac{\tilde{\mathcalz{f}}_{\, AD}(t)}{\mathcalz{f}_{\, ED}(t)} = \left(\frac{g_2(t)}{\mathcalz{f}_{\, AD}(t)} \right) \tilde{\mathcalz{f}}_{\, AD}(t) = g_2(t) \big( \, \mathcalz{d}_4(t)\lambda(L_M,L_P,s) \ + \ 1  \big) \ .
\end{equation} The sensitivity $s$, which is fixed in this study at $s=100\%$ (though should ideally be optimised), represents a universal and intuitive lever at the lender's disposal. Its adjustment can align with the lender's particular risk appetite and tolerances. At $s=0$, $g_3$ collapses back into $g_2$, though it purposefully resembles a more risk-adverse form of $g_2$ for $s>0$. Delinquency values are more varied than those of $g_2$ due to the inherent sensitivity to loan principals by design.